\documentclass[aps, prd, twocolumn, showpacs, floatfix, letterpaper,
 nofootinbib, superscriptaddress,longbibliography]{revtex4-2}

\usepackage{graphicx}
\usepackage{epsfig}
\usepackage{bm}
\usepackage{amsfonts}
\usepackage{pgf}
\usepackage{color}
\usepackage[T1]{fontenc}
\usepackage[latin1]{inputenc}
\usepackage{graphicx}
\usepackage[english]{babel}
\usepackage{amsmath}
\usepackage{amssymb}
\usepackage{amsfonts}
\usepackage{makecell}
\usepackage{hyperref}
\usepackage[draft,deletedmarkup=xout]{changes}
\usepackage{mathtools}
\usepackage{soul}
\usepackage{ulem}
\usepackage{relsize}
\usepackage{xcolor}
\usepackage{afterpage}

\definecolor{green}{rgb}{0.19,0.64,0.54}
\definecolor{blue}{rgb}{0,0,1}
\definecolor{reddish}{rgb}{0.65, 0.2, 0.2}
\definecolor{darkgreen}{rgb}{0.2,0.7,0.3}
\definecolor{darkblue}{rgb}{0.3,0.40,0.48}
\definecolor{gray}{rgb}{.8,.8,.8}

\hypersetup{,
	pdftitle={Coherent2},
  	pdfauthor={J. de Cabo Martins, P. Malkiewicz and P. Peter},
    colorlinks=true,    
    linkcolor=darkblue, 
    citecolor=darkgreen,
    filecolor=reddish,  
    urlcolor=blue,      
    linktocpage=true
}

\newcommand{\dd}{\mathrm{d}}
\newcommand{\ex}{\mathrm{e}}
\newcommand{\GN}{G_\textsc{n}}

\newcommand{\be}{\begin{equation}}
\newcommand{\ee}{\end{equation}}
\newcommand{\ba}{\begin{eqnarray}} 
\newcommand{\ea}{\end{eqnarray}}

\newcommand{\C}{\textsc{c}}
\newcommand{\F}{\textsc{f}}
\newcommand{\Ms}{\mathfrak{M}_\textsc{s}}
\newcommand{\Ns}{\mathfrak{N}_\textsc{s}}
\newcommand{\Ts}{\mathfrak{T}_\textsc{s}}

\newcommand{\K}{\mathfrak{K}}
\newcommand{\Ls}{\mathfrak{L}_\textsc{s}}

\newcommand{\chiC}{\frac{3(1-w)}{1-3w}}

\newcommand{\JM}{\textcolor{darkgreen}}

\begin{document}

\title{Ambiguous power spectrum from a quantum bounce}

\author{Jaime de Cabo Martin}
\email{jaime.decabomartin@ncbj.gov.pl}
\affiliation{National Centre for Nuclear Research, Pasteura 7, 02-093
Warszawa, Poland}

\author{Przemys{\l}aw Ma{\l}kiewicz}
\email{Przemyslaw.Malkiewicz@ncbj.gov.pl}
\affiliation{National Centre for Nuclear Research, Pasteura 7, 02-093
Warszawa, Poland}

\author{Patrick Peter}
\email{peter@iap.fr}

\affiliation{${\cal G}\mathbb{R}\varepsilon\mathbb{C}{\cal O}$ --
Institut d'Astrophysique de Paris, CNRS \& Sorbonne Universit\'e, UMR
7095 98 bis boulevard Arago, 75014 Paris, France}

\affiliation{Centre for Theoretical Cosmology, Department of Applied
Mathematics and Theoretical Physics, University of Cambridge,
Wilberforce Road, Cambridge CB3 0WA, United Kingdom}

\begin{abstract}
A quantum cosmological bouncing model may exhibit an
ambiguity stemming from the nonclassical nature of the
background evolution: two classically equivalent theories
can produce two qualitatively different potentials sourcing
the perturbations. We derive explicitly the quantum
canonical transformation involving the quantum background to
show how it leads to inequivalent theories. We identify the
relevant quantum parameter describing the difference and
expand upon the ambiguity by calculating the expected power
spectra produced for initial quantum vacuum fluctuations in
the contracting phase of both potentials. We find that one
spectral index corresponds to all values of this parameter
but one, while the other thus represents a set of measure
zero.
\end{abstract}

\maketitle

\section{Introduction}

In a previous work~\cite{Martin:2021dbz}, we studied a
cosmological model whose dynamics is led by general
relativity (GR) and a perfect fluid with arbitrary equation
of state. Classically, the solutions either contract toward
or expand from a singularity: quantization permits to
regularize the trajectories (dubbed ``semiquantum'' in
\cite{Martin:2021dbz}), thereby leading to a quantum
bouncing behaviour.

Quantization of a cosmological model replaces the
four-dimensional spacetime with what we shall call a
``quantum spacetime'' that violates the properties of
classical geometry. For instance, the dynamical law for a
field in a classical spacetime can be formulated in terms of
different field variables. In the Hamiltonian formalism used
in this work, these variables are connected by canonical
transformations.  The fact that they all undergo physically
equivalent evolutions can be viewed as a manifestation of a
unique underlying background spacetime that imposes a unique
dynamical law on the field. In passing to quantum theory of
the background spacetime, on the other hand, the use of
different field variables may result in their very different
quantum dynamics. It is so because the field does no longer
propagate in a fixed spacetime but rather in a ``quantum
spacetime'', and the physical discrepancies between the
evolutions of different field variables reflect the
``quantumness of spacetime''. As found in
Ref.~\cite{Martin:2021dbz}, simple re-scalings of the
curvature perturbation in a quantum Friedmann universe by
powers of the scale factor produce different gravitational
potentials in the Mukhanov-Sasaki equation, thereby making
the dynamics of the perturbation depend on the choice of the
field variable employed in its quantization.

In the present work we study the physical consequences of
the ambiguity in the dynamical law for the scalar
perturbation due to quantization of the background
spacetime. If the infinitely many gravitational potentials
found in \cite{Martin:2021dbz} actually correspond to
infinitely many physical predictions, then the theory is to
be considered unphysical. Currently, the literature (see
e.g. \cite{Peter:2006id, Peter:2006hx, Martin2004a,
Malkiewicz:2020fvy}) provides only one solution to the
primordial power spectrum from a quantum bounce. We note
that the simplest known quantum bounces produce only
blue-tilted power spectrum contrary not only to inflationary
predictions but also to observations; finding new solutions
could improve this situation.

Our work also pertains to the question of the usefulness of
the Mukhanov variable for describing scalar perturbations in
a quantum universe. In inflationary models based on
classical backgrounds it is convenient to use the Mukhanov
variable because it allows to asymptotically define a
quantum vacuum in the same way as for flat spacetime.
However, it is not the only valid choice for the
perturbation variables even in the context of inflation as
discussed, e.g.,  in Ref.~\cite{Grain:2019vnq}. In a fully
quantum universe, the issue is even less clear as the choice
of perturbation variables can influence both the definition
of the initial vacuum state and the dynamics of
perturbations. Thus, in the context of our work, it is
natural to ask whether the Mukhanov variable remains a
preferred choice in a fuller, more quantum description of
the primordial universe or if it should be replaced with
another, better-suited, variable. The latter situation would
not be exceptional as there exist situations for which the
Mukhanov variable may become singular and other variables
as, e.g,. the Bardeen potential, must be used instead.

As the last point of this introduction, let us remark that
we view our framework of quantum fields in quantum spacetime
as a truncation of a full theory of quantum gravity. The
dynamics in the latter would naturally be expressed in some
internal time variable. Even if many internal time variables
are available, the particular choice one makes does not seem
to be crucial for the physical predictions of quantum
gravity (we refer the interested reader to some of our
previous works devoted to this issue
\cite{Malkiewicz:2015fqa, Malkiewicz:2016hjr,
Malkiewicz:2017cuw, Malkiewicz:2019azw,Malkiewicz:2022szx}).
Nevertheless, if a truncation is to be consistent, it should
make use of a unique internal time variable for quantizing
and describing all dynamical variables, both for the
background and the perturbations. We emphasize that our
framework satisfies this requirement.

The plan of this work is as follows. In Sec.
\ref{ambiguity}, we introduce our cosmological model: we
first define the classical model with a special attention
paid to the definition of perturbation variables, then we
quantize it and set up its semi-classical approximation. We
next identify and discuss the dynamical ambiguity in the
resulting quantum model. In Sec. \ref{numerical} and
\ref{tale}, we study the ambiguity in detail: we solve the
dynamical equations and find all the possible physical
predictions for the amplitude and the spectral index of
primordial perturbations, which could be derived from the
model. We find that there are only two types of predictions
that can be produced by quantum bounce models.
Sec.~\ref{conclusion} gathers our conclusions.

\section{The ambiguity}
\label{ambiguity}

Before going on to the new results obtained in this work,
let us summarize the framework of the semiquantum ambiguity.

\subsection{Background dynamics}

We begin by defining the model used throughout. We expand
our spacetime manifold $\mathcal{M}$ as $\mathcal{M} \simeq
\mathbb{R} \times \mathbb{T}^3$, whose background metric is
given by the isotropic and homogeneous flat
Friedmann-Lema\^{\i}tre-Robertson-Walker (FLRW) metric
\begin{equation}
\dd s^2 = -N^2(\tau)\dd\tau^2 + a^2(\tau) \gamma_{ij}\dd x^i
\dd x^j, \label{FLRW}
\end{equation}
in units in which the velocity of light is $c=1$. We assume
that the spatial part $\mathbb{T}^3$ is compact with
coordinate volume
\begin{equation}
\mathcal{V}_0 := \int_{\mathbb{T}^3} \sqrt{\gamma} \dd^3 x,
\label{Vo}
\end{equation}
while the scale factor $a(\tau)$ has its dynamics driven by
GR sourced by a perfect fluid with constant equation of
state parameter $0<w=p/\rho<1$ (with $p$ the pressure and
$\rho$ the energy density). The overall
Einstein-Hilbert-Schutz action then
reads~\cite{Schutz:1970my, Schutz:1971ac}
\begin{align}\begin{split}
\mathcal{S}_\textsc{ehs}=\frac{1}{2\kappa}\int \dd^4 x
  \sqrt{-g} R+\int \dd^4 x
  \sqrt{-g} P(w,\phi),\end{split}
\end{align}
where $\kappa = 8\pi\GN$ and $\phi$ defines the cosmic fluid
flow. We use the expansion of this action to second order
and an adaptation of the fully canonical formalism
from~\cite{Malkiewicz:2018ohk} so as to obtain a Hamiltonian
description in which only the truly physical degrees of
freedom are considered.

\subsubsection{Classical dynamics}

Setting the lapse function to $N=(1+w)a^{3w}$, the fluid
part of the action contributes a linear momentum term to the
Hamiltonian~\cite{Bergeron:2013jva}, so the fluid merely
serves as a clock in what follows. Going from $\{ a, p_a\}$
to $\{ q, p\}$, defined by
\begin{equation}
q=\frac{4\sqrt{6}}{3(1-w)\sqrt{1+w}}a^{\frac{3}{2}(1-w)}
\equiv \gamma a^{\frac{3}{2}(1-w)},
\label{defGamma}
\end{equation}
and 
\begin{equation}
p=\frac{\sqrt{6(1+w)}}{2\kappa_0} a^{\frac{3}{2}(1+w)}H,
\end{equation}
with $H=\dot{a}/(Na)$ the Hubble rate and
$\kappa_0=\kappa/\mathcal{V}_0$, the zeroth-order
gravitational Hamiltonian reads
\begin{equation}
H^{(0)} = 2\kappa_0 p^2,
\label{h0}
\end{equation}
where we reverted the direction of time with respect to the
fluid variable to make it positive. We hereafter use $\tau$
as internal clock, assuming it coincides with the FLRW time
set in \eqref{FLRW}.

The solutions of the equations of motion stemming from
\eqref{h0} read
\begin{equation}
q_\text{cl}(\tau)=4\sqrt{2\kappa_0
H^{(0)}}(\tau-\tau_\text{s})\quad \text{and} \quad
p_\text{cl}(\tau)=\sqrt{\frac{H^{(0)}}{2\kappa_0}},
\label{qpSol}
\end{equation}
where $H^{(0)}$ is a constant. These classical trajectories
either terminate at or emerge from the singularity $q\to 0$
(and thus $a\to 0$) at time $\tau_\text{s}$. They are
straight lines in phase space $\{ q, p\}$ with constant
$p$~\cite{Malkiewicz:2015fqa}.

\subsubsection{Quantum dynamics}

Quantization of \eqref{h0} is not as trivial as it seems. Of
course, one can merely impose a canonical rule and
straightforwardly set $p\mapsto \widehat{P} = -i\hbar
\partial/\partial q$, but it turns out to be more general to
make the replacement
\begin{equation}\label{qh0}
H^{(0)}\mapsto \widehat{H}^{(0)}=2\kappa_0\left(\widehat{P}^2+
{\mathfrak{c}_{0}}\widehat{Q}^{-2}\right),
\end{equation}
thereby introducing an arbitrary parameter
$\mathfrak{c}_{0}\geqslant 0$ which not only can account for
other quantization procedures, such as making use of the
affine group as the symmetry of
quantization~\cite{Bergeron:2013jva,Malkiewicz:2019azw}, but
also provides a means of parametrizing the factor ordering
ambiguity. For $\mathfrak{c}_{0}=0$, one recovers the
canonical case, while for $\mathfrak{c}_{0}\geqslant
\frac{3}{4}\hbar^2$, the Hamiltonian $\widehat{H}^{(0)}$
becomes essentially self-adjoint and no boundary condition
is required at $q=0$ to ensure a unique and unitary
dynamics.

A useful approximate semiclassical (or, rather,
semiquantum~\cite{Martin:2021dbz}) solution is obtained with
a family of coherent states~\cite{Klauder:2015ifa} built
upon a fiducial state $|\xi\rangle$, satisfying
$\langle\xi|\widehat{Q}|\xi\rangle =1$ and
$\langle\xi|\widehat{P}|\xi\rangle =0$, namely
\begin{equation}
|q(\tau),p(\tau)\rangle = \ex^{i p(\tau)
\widehat{Q}/\hbar}\ex^{-i \ln q(\tau)
\widehat{D}/\hbar}|\xi\rangle, \label{CohSta}
\end{equation}
in which the operator $\widehat{D}$, called dilation, is
defined by $\widehat{D} = \frac{1}{2} (\widehat{Q}
\widehat{P} + \widehat{P} \widehat{Q})$. These states,
written in short $|q,p\rangle$ when there is no risk of
confusion, are such that the expectation values of
$\widehat{P}$ and $\widehat{Q}$ are respectively $\langle
q,p|\widehat{Q}|q,p \rangle = q(\tau)$ and $\langle
q,p|\widehat{P}|q,p \rangle = p(\tau)$.

Setting
\begin{equation}
 H_\text{sem} =\langle q,p|\widehat{H}^{(0)}|q,p\rangle =
 2\kappa_0\left(p^2+\frac{\mathfrak{K}}{q^2}\right),
 \label{Hsem}
\end{equation}
with $\mathfrak{K}>0$ is a constant depending on the family
of coherent states chosen for the semiquantum approximation,
it can be shown that $q(\tau)$ and $p(\tau)$ satisfy the
Hamilton equations derived from $H_\text{sem}$, with
solutions
\begin{subequations}
\label{solsem}
\begin{align}
  &q=q_\textsc{b} \sqrt{1+(\omega\tau)^2}, \\
  p&= \frac{q_\textsc{b} \omega^2}{4\kappa_0}
  \frac{\tau}{\sqrt{1+(\omega\tau)^2}},
\end{align}
\end{subequations}
where $q_\textsc{b}^2 = 2 \kappa_0
\mathfrak{K}/H_\text{sem}$ and $\omega^2 = 4
H^2_\text{sem}/\mathfrak{K}$. The singularity is clearly
replaced by a bouncing behaviour, as expected.

\subsection{Perturbations}

Compactness of space implies discrete Fourier wavevectors
$\bm{k}$ and the second-order Hamiltonian $H^{(2)}$ is
naturally a sum over independent contributions
$H^{(2)}_{\bm{k}}$ for each mode. The relevant perturbation
variables, noted $\phi_{\bm{k}}$ in what follows, are
obtained as a combination of the fluid variable
$\delta\phi_{\bm{k}}$ and the intrinsic curvature
perturbation $\delta R_{\bm{k}}$, namely
\begin{widetext}
\begin{align}
\begin{split}
\phi_{\bm{k}} & = \frac{1}{\sqrt{2\kappa_0}\mathcal{V}_0 }
\left[
\frac{\bar{p}^{\phi}\delta\phi_{\bm{k}}}{\sqrt{w(1+w)}p} +
\sqrt{\frac{6}{w}} a^{\frac32 (1-w)}
\frac{a^2}{4\mathcal{V}_0^{2/3}k^2} \delta
R_{\bm{k}}\right],\\ \pi_{\phi,{\bm{k}}} & =
\sqrt{2\kappa_0} \left[ \sqrt{w(1+w)}
\frac{a^{3(1+w)}p}{|\bar{p}^{\phi}|^{1+w}}
\delta\rho_{\bm{k}} - \frac{3 (1-w)
a^{2}p}{8\mathcal{V}_0^{2/3}k^2}  \sqrt{\frac{1+w}{w}}
\delta R_{\bm{k}} -\sqrt{\frac{3}{8w}} (1+w) a^{-\frac32
(1-w)}~\bar{p}^{\phi}\delta\phi_{\bm{k}}\right],
\label{DefPhiPi}
\end{split}
\end{align}
\end{widetext}
where $\delta\rho_{\bm{k}}$ represents the fluid energy
density perturbation, while the background variable
$\bar{\phi}$ and its momentum $\bar{p}^{\phi}$ describe the
fluid, with $-i\bm{k}\bar{p}^{\phi}\delta\phi_{\bm{k}}$
defining the flow of the fluid energy through the surface
orthogonal to $\bm{k}$ (see details in
Ref.~\cite{Malkiewicz:2018ohk}).

In term of the above variables, the Hamiltonian reads
\begin{equation}
H^{(2)}_{\bm{k}}=
\frac{1}{2}\left[ \left| \pi_{\phi,{\bm{k}}}\right|^2
+ w(1+w)^2 \left(\frac{q}{\gamma}\right)^{4 r_\F}
k^2 \left| \phi_{\bm{k}}\right|^2\right],
\label{HAMpert}
\end{equation}
with
\begin{equation}
r_\F=-\frac{1-3w}{3(1-w)},
\label{rF}
\end{equation}
whose numerical value lies in the range $-\frac13\leq
r_\F\leq 0$ for the range of equation of state $0\leq w\leq
\frac13$ we are concerned with.

Curvature perturbations, which are observationally relevant
gauge-invariant variables, can be easily derived from
\eqref{DefPhiPi}. They are often calculated on comoving
hypersurfaces, thereby defining $\mathcal{R}_{\bm{k}}$
through $a^2 k^2 \mathcal{R}_{\bm{k}} =- 4 \delta
R_{\bm{k}}\big|_{\delta\phi=0}$ (with a conventional minus
sign), reading
\begin{align}
\mathcal{R}_{\bm{k}} = - \sqrt{\frac{w\kappa_0}{3}}
a^{-\frac32 (1-w)}\phi_{\bm{k}},
\end{align}
or uniform-density hypersurfaces ($\zeta_{\bm{k}}$), i.e.
$a^2 k^2 \zeta_{\bm{k}} = 4 \delta
R_{\bm{k}}\big|_{\delta\rho=0}$, which is
\begin{align}
\zeta_{\bm{k}} = \frac{\mathcal{V}_0}{2} \sqrt{\frac{\kappa_0}{3w}}
\frac{ (1+w)\phi_{\bm{k}}}{a^{\frac12 (1-w)}}  +
\frac{\pi_{\phi,{\bm{k}}} }{3 p \sqrt{2w(1+w)\kappa_0}}.
\end{align}
Note that it is usually assumed, e.g. using initial
conditions for the expanding Universe at the end of
inflation, that $\phi_{\bm{k}} \gg \pi_{\phi,{\bm{k}}}$,
leaving only the first term.

The classical theory based on the Hamiltonian
\eqref{HAMpert} can be written using a different set of
variables by performing a time-dependent canonical
transformation. Let us define new variables via the
rescaling
\begin{equation}
v_{\bm{k}}=Z \phi_{\bm{k}} \qquad \hbox{and} \qquad
\pi_{v,{\bm{k}}} = Z^{-1}\pi_{\phi,{\bm{k}}}+
\frac{\dot{Z}}{Z^2}\phi_{\bm{k}},
\label{ntoms}
\end{equation}
where $Z$ is any nonvanishing phase space function. This
transforms the Hamiltonian into
\begin{equation}
H^{(2)}_{\bm{k}}\! = \frac{1}{2} Z^2 \left\{
\left| \pi_{v,{\bm{k}}}\right|^2
\! +\! \left[ \frac{w(1+w)^2}{Z^4}
\left(\frac{q}{\gamma}\right)^{4r_\F}
\! \! k^2 - \mathcal{V}\right] \left| v_{\bm{k}}\right|^2
\right\},
\label{H2k}
\end{equation}
where the potential $\mathcal{V}$ is defined through
\begin{equation}
\mathcal{V} = \frac{1}{Z^4}
\left[ \frac{\ddot{Z}}{Z} -2 \left(
  \frac{\dot{Z}}{Z} \right)^2 \right].
\label{calV}
\end{equation}
An interesting and useful one-parameter family of such scale
transformations is obtained by setting
\begin{equation}
Z_s = \sqrt{1+w}\left(\frac{q}{\gamma}\right)^s,
\label{Zs}
\end{equation}
which yields
\begin{equation}
H^{(2)}_{\bm{k}} = \frac{1+w}{2} \left(\frac{q}{\gamma}\right)^{2s}
\left( \left| \pi_{v,{\bm{k}}}\right|^2
+ \Omega_s^2 \left| v_{\bm{k}}\right|^2 \right),
\label{T1}
\end{equation}
with 
\begin{equation}
\Omega_s^2 = \left(\frac{q}{\gamma}\right)^{4(r_\F-s)} w k^2
+ \frac{s(s+1)(4\kappa_0)^2}{q^2 (1+w)^2}
\left(\frac{q}{\gamma}\right)^{-4s} p^2,
\label{T2}
\end{equation}
where we used the classical solution \eqref{qpSol}. It
should be noted at this point that setting $s=r_\F$ and
introducing the conformal time $\eta$ through
\begin{equation}
\dd\eta=(1+w)\left( \frac{q}{\gamma}\right)^{2r_\F} \dd\tau
=  Z_{r_\F}^2 \dd\tau
\label{eta}
\end{equation}
removes the overall coefficient in front of the Hamiltonian
\eqref{T1}, turning the kinetic term into its canonical
form, while \eqref{T2} takes the simple and usual form
$$
\Omega_{r_\F}^2 = w k^2 - \frac{Z_{r_\F}''}{Z_{r_\F}},
$$
where a prime means a derivative with respect to the
conformal time $\eta$. The variable $v_{\bm{k}}$ in this
case is nothing but the well-known Mukhanov-Sasaki variable:
its usual definition $v_{\bm{k}} = \sqrt{1+w} a^{-\frac12
(1-3w)} \phi_{\bm{k}}$ indeed becomes
\begin{equation}
v_{\bm{k}} = - \sqrt{\frac{3(w+1)}{w\kappa_0}}a
\mathcal{R}_{\bm{k}} ,
\label{vR}
\end{equation}
so that the function $v/a$, whose spectrum we compute below,
can be identified, up to an irrelevant constant, with the
curvature perturbation on the comoving hypersurfaces.

\begin{figure}
\includegraphics[width=0.45\textwidth]{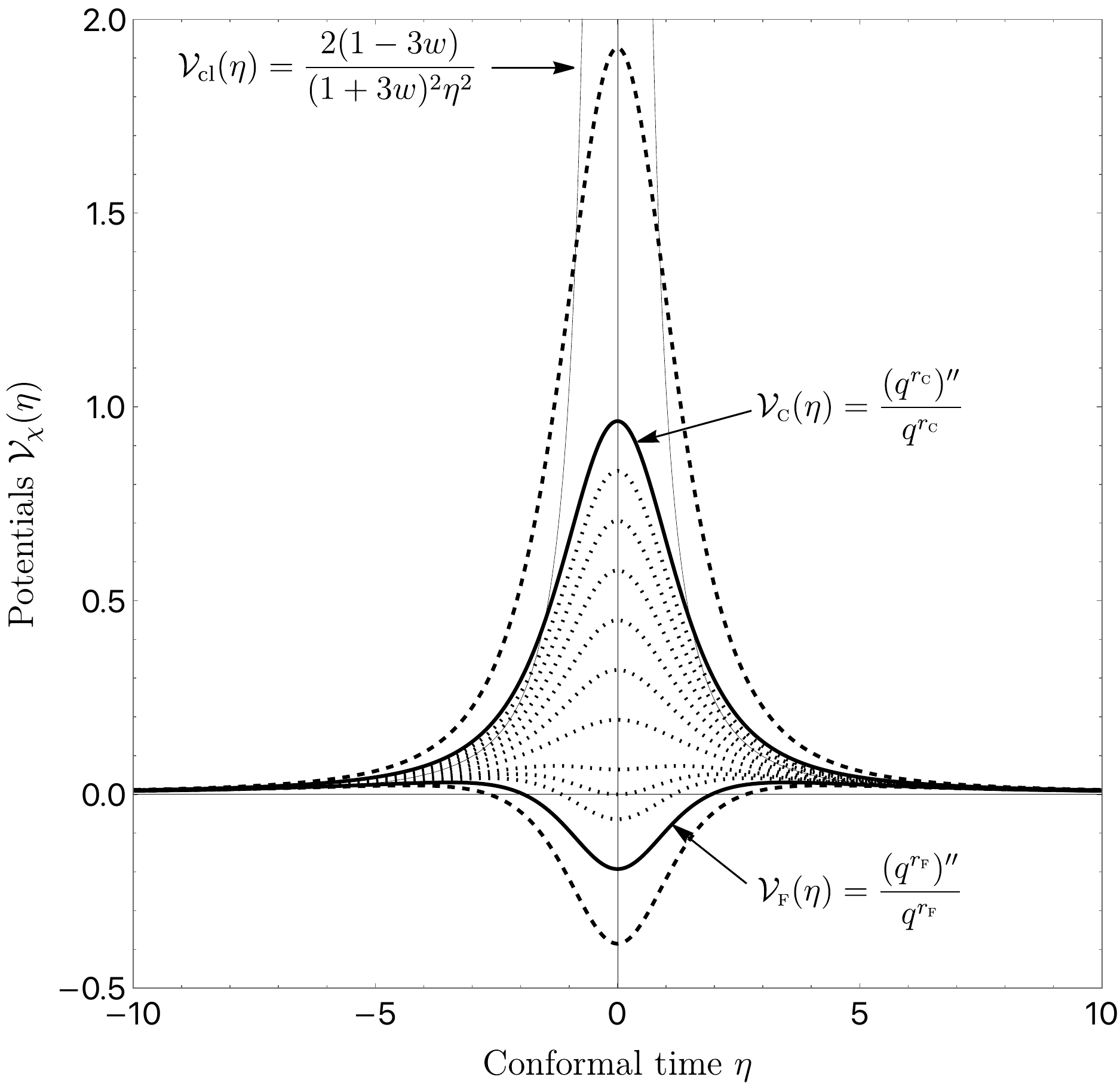}

\caption{The potential $\mathcal{V}_\chi$ sourcing the
perturbations for $w=0.2$ and different values of the
phenomenological parameter $\chi$. They all converge far
from the bounce where they behave as $\mathcal{V}_\text{cl}$
[Eq.~\eqref{Vcl}] shown as a thin line. The two extreme with
$\chi=\chi_\F (w)$ and $\chi=\chi_\C(w)$ are shown as full
thick lines, while the dotted curves represent a selection
of values between these extreme. The dashed curves are for
$\chi < \chi_\F$ and $\chi > \chi_\C$.}

\label{FigPots}
\end{figure} 

One can introduce the quantization of the background
following a two step procedure: one first replaces the
zeroth order quantities by the corresponding operators using
the rules
\begin{equation}\begin{split}
q^{\alpha} & \mapsto{\mathfrak{l}(\alpha)}
\widehat{Q}^{\alpha},\\ q^{\alpha}p^2 & \mapsto
\mathfrak{a}(\alpha) \widehat{Q}^{\alpha}\widehat{P}^2
-i\mathfrak{a}(\alpha)\alpha\widehat{Q}^{\alpha-1}\widehat{P}+ \mathfrak{c}(\alpha) \widehat{Q}^{\alpha-2}, \label{ClassQsym}
\end{split}
\end{equation}
in which $\mathfrak{l}(\alpha)$, $\mathfrak{a}(\alpha)$ and
$\mathfrak{c}(\alpha)$ are positive, $\mathfrak{a}(\alpha)$
being dimensionless and $\mathfrak{c}(\alpha)\propto
\hbar^2$. Details of this procedure are to be found in a
dedicated appendix of Ref.~\cite{Martin:2021dbz}. From now
on, we move to natural units in which $\hbar =1$. Combining
this step with the usual quantization of the perturbation
variables $v_{\bm{k}}$ and $\pi_{v,{\bm{k}}}$, maps the
classical Hamiltonian \eqref{T1} into a quantum one:
\begin{equation}
\widehat{H}^{(2)}_{\bm{k}} = \frac{1+w}{2}~\mathfrak{l}(2s)
\left(\frac{\widehat{Q}}{\gamma}\right)^{2s} \left( \left|
\widehat{\pi}_{v,{\bm{k}}}\right|^2 + \widehat{\Omega}_s^2
\left| \widehat{v}_{\bm{k}}\right|^2 \right), \label{T1Q}
\end{equation}
with
\begin{widetext}
\begin{align}
\widehat{\Omega}_s^2 =
\frac{\mathfrak{l}(4r_\F-2s)}{\mathfrak{l}(2s)}
\left(\frac{\widehat{Q}}{\gamma}\right)^{4(r_\F-s)} w k^2 +
\frac{s(s+1)(4\kappa_0)^2}{(1+w)^2} g_1(s)
\left(\frac{\widehat{Q}}{\gamma}\right)^{-4s}
\widehat{Q}^{-2}\left[ \widehat{P}^2 +2 i
(s+1)\widehat{Q}^{-1}\widehat{P}+ g_2(s) \widehat{Q}^{-2}
\right], \label{T2Q}
\end{align}
\end{widetext}
where the function
$g_1(s)=\mathfrak{a}(-2s-2)/\mathfrak{l}(2s)$ can be set to
unity by conveniently choosing the fiducial vector, while
$g_2(s) = \mathfrak{c}(-2s-2)/\mathfrak{a}(-2s-2)$ is
unknown at this stage.

The second step consists in averaging the corresponding
operator (function of $\widehat{Q}$ and $\widehat{P}$) in
the semiquantum state $|q(\tau),p(\tau)\rangle$, leading to
a Hamiltonian depending only on the perturbation variables,
the background being then described by a trajectory. In
practice, we use the definition
\begin{equation}
H^s_{\bm{k}}  (\eta) := \langle q(\eta),p(\eta)|
\widehat{H}^{(2)}_{\bm{k}}| q(\eta),p(\eta) \rangle,
\label{HC}
\end{equation}
and assume the parametric phase-space trajectory $\{
q(\tau),p(\tau)\}$ to be given by the solutions
\eqref{solsem}.

Applying to these classically equivalent formulations the
2-step quantization procedure discussed above, and upon
canonical quantization of the variable $v_{\bm{k}}$, one
finds
\begin{equation}
H^s_{\bm{k}} (\eta) = \frac{\Ms{(1+w)}}{2} 
\left(\frac{q}{\gamma}\right)^{2s}
\left( \left| \widehat{\pi}_{v,{\bm{k}}} \right|^2
+ \check{\Omega}_s^2 \left| \widehat{v}_{\bm{k}}\right|^2
\right),
\label{MSqhHsem}
\end{equation}
where
\begin{align}
\check{\Omega}_s^2 = & \Ls\Ms^{-1} 
\left(\frac{q}{\gamma}\right)^{4(r_\F-s)} w k^2
\nonumber \\
& + \frac{\Ms^{-1}s(s+1)(4\kappa_0)^2}{q^{2}(1+w)^2}
\left(\frac{q}{\gamma}\right)^{-4s}
\left( \Ns p^2 + \frac{ \Ts}{q^2} \right),
\label{Omegav}
\end{align}
and $\Ls$, $\Ms$, $\Ns$, $\Ts$ are four arbitrary parameters
(see again Ref.~\cite{Martin:2021dbz} for details).

Expanding $\widehat{v}_k$ in creation and annihilation
operators yields the evolution for the mode functions,
denoted $v^s_k (\tau)$, which only depend on the amplitude
$k=|\bm{k}|$ of the wavevector and not the direction
$\bm{k}/k$. All the above models can be easily compared
provided one rescales the respective mode functions $v^s_k
(\tau)$ to a common variable, the Mukhanov-Sasaki variable
introduced above and denoted by $\tilde{v}^s_k (\tau)$. Upon
rescaling the mode function $v^z_k (\tau)$ via the scale
transformation \eqref{T1} with $Z_{r_\F-s}\propto
\left(q/\gamma\right)^{r_\F-s}$ based on the semiclassical
equations of motion,
\begin{equation}
\tilde{v}^s_k = Z_{r_\F-s} v^s_k,
\end{equation}
and using the conformal time defined through Eq.~\eqref{eta}
above, one obtains the respective $s-$dependent
Mukhanov-Sasaki equation for the mode functions
\begin{equation}
\frac{\dd \tilde{v}^s_k}{\dd\eta^2} + \left( k_\text{eff}^2 -
\mathcal{V}_s\right) \tilde{v}^s_k = 0, \label{d2vs}
\end{equation}
where the effective wave number is $k_\text{eff} = k \sqrt{w
\Ls\Ms^{-1}}$ and the potential reads
\begin{align}
\mathcal{V}_s = \frac{(4\kappa_0)^2}{\Ms (1+w)^2 q^2}
\left(\frac{q}{\gamma}\right)^{-4r_\F} \left[\mathcal{A}(s) p^2 +
\frac{\mathcal{B}(s)}{q^2} \right], \label{Vs}
\end{align}
with $\mathcal{A}(s)= - \Ns s (s+1) - \Ms^{-1}
\left(r_\F-s\right) \left(r_\F-s+1\right)$ and
$\mathcal{B}(s) = - \Ts s (s+1) - \Ms^{-1} \left(r_\F-s
\right) \left(2s - \mathfrak{K}\right)$. In the large $q$
limit, one must recover the classical limit, which demands
that $\mathcal{A}(s)>0$, a constraint we assume from now on.
We are thus left with two possible cases depending on the
sign of $\mathcal{B}(s)$. Absorbing the irrelevant constants
into redefinitions of time and wavenumbers, we find that the
potentials for the special representative cases $s=0$ and
$s=r_\F$ can be transformed into
\begin{subequations}
\begin{align}
&\mathcal{V}_0 =
\frac{8(2\kappa_0)^2(1-3w)}{9(1-w^2)^2 q^2 }
\left(\frac{q}{\gamma}\right)^{-4r_\F}
\left[p^2-\frac{3(1-w) \mathfrak{K}}{2q^2}\right],
\label{Vf}\\
&\mathcal{V}_{r_\F} = 
\frac{8(2\kappa_0)^2(1-3w)}{ 9(1-w^2)^2q^2}
\left(\frac{q}{\gamma}\right)^{-4r_\F}
\left(p^2+\frac{{\Ts}/{\Ns}}{q^2}\right),
\label{Vc}
\end{align}
\label{pots}
\end{subequations}
displayed in Fig.~\ref{FigPots}. They differ in the
numerical coefficient of the last term $\propto q^{-2}$,
namely $\Ts/\Ns>0$ in \eqref{Vc} and $-\frac32(1-w)
\mathfrak{K} <0$ in \eqref{Vf} as $w<1$. Setting this
coefficient to $\chi \mathfrak{K}$, thereby defining $\chi$,
we write the generic potential as
\begin{equation}
\mathcal{V}_\chi = 
\frac{8(2\kappa_0)^2(1-3w)}{9(1-w^2)^2q^2}
\left(\frac{q}{\gamma}\right)^{-4r_\F}
\left(p^2 +
\frac{\chi \mathfrak{K}}{q^2}\right),
\label{VchiK}
\end{equation}
which, upon using the background solution \eqref{solsem} and
expressing $\mathfrak{K}$ and $H_\mathrm{sem}$ in terms of
$q_\textsc{b}$, $\omega$ and $\kappa_0$, namely
$$\mathfrak{K}=\frac{q_\textsc{b}^4 \omega^2}{16\kappa_0^2}
\ \ \ \hbox{and}\ \ \ H_\mathrm{sem} =
\frac{q_\textsc{b}^2 \omega^2}{8\kappa_0},$$
becomes
\begin{equation}
\mathcal{V}_\chi = \frac{2\omega^2}{9 Z_{r_\F}^4}
\frac{1-3w}{(1-w)^2} \frac{\chi + (\omega
\tau)^2}{\left[1+(\omega \tau)^2 \right]^2},
\label{Vchi}
\end{equation}
with $Z_{r_\F}$ given by \eqref{Zs}(with $s\to r_\F$). In
what follows, we shall accordingly note $\chi =
\Ts/(\mathfrak{K}\Ns)$ and $\chi_\F = -\frac32(1-w)$.

As discussed in Ref.~\cite{Martin:2021dbz}, there are two
special values for $\chi$, namely $\chi = \chi_\F$ for which
the potential is $\mathcal{V}_\F = (q^{r_\F})''/q^{r_\F}$,
and $\chi = \chi_\C = 3(1-w)/(1-3w)$, leading to
$\mathcal{V}_\C = (q^{r_\C})''/q^{r_\C}$, with
\begin{equation}
r_\C = 1+ r_\F = \frac{2}{3(1-w)},
\label{rC}
\end{equation}
implying $\frac23\leq r_\C \leq 1$ with $0\leq w\leq
\frac13$. It can be argued that these values, naturally
reproducing the degenerate classical case
\begin{equation}
\mathcal{V}_\text{cl} =
\frac{(q_\text{cl}^{r_\F})''}{q_\text{cl}^{r_\F}} =
\frac{(q_\text{cl}^{r_\C})''}{q_\text{cl}^{r_\C}} =
\frac{2(1-3w)}{(1+3w)^2 \eta^2}, \label{Vcl}
\end{equation}
with $q_\text{cl}$ and $p_\text{cl}$ the classical solutions
\eqref{qpSol}, are the only physically acceptable ones. This
argument is reinforced by the fact that $\chi_\F$ and
$\chi_\C$ imply the potential to depend only on the
background quantities $w$ and $\mathfrak{K}$, in agreement
with the idea of perturbation theory. We shall however in
what follows assume that $\chi$ is an arbitrary real
parameter and restrict attention to these particular cases
whenever necessary. Fig~\ref{FigPots} shows such possible
cases, including the situations $\chi < \chi_\F = -1/r_\C$
and $\chi > \chi_\C= -1/r_\F$, illustrating that these
extreme points do not lead to anything particular apart from
the fact that they permit a simple writing of the potential.
In fact, in the large time limit $\omega\tau\gg 1$, the term
in $\chi$ in \eqref{Vchi} is in any way negligible and it
can be checked explicitly that the classical case
\eqref{Vcl} is also recovered for any value of $\chi$.

One can also note that the potential $\mathcal{V}_\chi$ in
\eqref{Vchi} can be given a simple form for an arbitrary
$\chi$, namely
\begin{equation}
\mathcal{V}_\chi  = \alpha_\chi^2
\frac{\left(q^{r_\chi}\right)''}{q^{r_\chi}},
\label{qrdd}
\end{equation}
with
$$
\alpha_\chi^2 = \frac{2(1-3w)\chi^2}{9(1-w)^2 \left[
1+\left(1+2r_\F\right)\chi\right]}
$$
and
$$
r_\chi = \frac{1}{\chi} +r_\F+ r_\C.
$$
One can easily check that for $w\in ]-\frac13,\frac13[$,
$\alpha_\chi^2$ is indeed positive definite, varying from
$\alpha_{\chi_\F} =1$ to $\alpha_{\chi_\C} =1$ with
$\alpha_0 =0$. In what follows, we assume $r_\F \leq r \leq
r_\C$, which is equivalent to $\chi_\F \leq \chi \leq
\chi_\C$.

The ambiguity is clear from Fig~\ref{FigPots}: the parameter
$\chi$ stems from the arbitrary choice that is made among
various classically equivalent theories to quantize. Once
this choice is made, and the factor ordering is taken care
of, the potential for the perturbations is completely fixed,
but it depends on the actual value of $\chi$, which is
\emph{a priori} not fixed by any physical principle. The
resulting spectrum, to which we now turn, therefore also
depends on this unphysical parameter, thereby leading to
ambiguous physical predictions.

A more detailed examination of Eqs.~\eqref{Vc} and
\eqref{Vf} however shows that the ambiguity is twofold.
Using the fluid parametrization yields $\chi = \chi_\F < 0$,
which is entirely fixed once the barotropic index of the
equation of state $w$ and the background solution
\eqref{solsem} are known. On the other hand, the conformal
case depends on $\chi=\Ts/{(\K\Ns)}$, which is not fixed by
the background but rather by the quantization procedure,
with the special case $\chi=\chi_\C=\chiC>0$ again solely
fixed by the equation of state. Therefore, the first
question to address concerns the sign of $\chi$ and, for
positive values, its numerical value.

\begin{figure}
\includegraphics[width=0.48\textwidth]{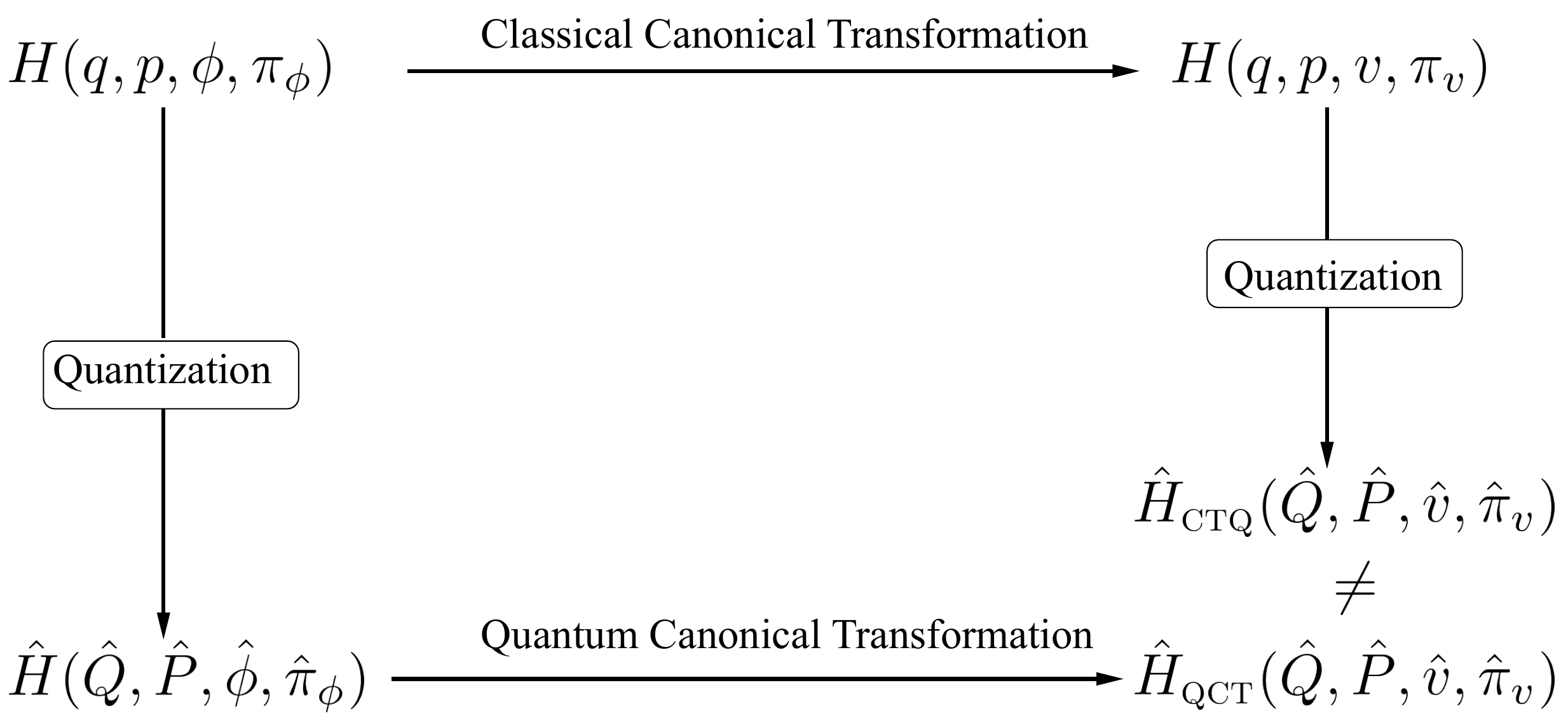}

\caption{The origin of our ambiguity explained by the
process through which one changes variables: starting with a
given classical first order theory with Hamiltonian
$H(q,p,\phi,\pi_\phi)$ depending on background $(q,p)$ and
perturbation $(\phi,\pi_\phi)$ variables, one can either
perform a classical canonical transformation and quantize
the resulting theory, or first quantize the original theory
and subsequently perform a quantum canonical transformation.
These operations in general lead to different quantum
theories.}

\label{QclCT}
\end{figure} 

\subsection{Discussion}
\label{discussion}

Let us conclude the present section by emphasizing the
quantum origin of our new ambiguity.

First, we introduced a set of different basic variables for
the perturbation field through Eq.~\eqref{ntoms}. They all
represent equally valid choices, their dynamics being, as
expected, generated by different Hamiltonians \eqref{H2k}
that involve different background observables for coupling
the perturbation variables to the background spacetime.
Next, we quantized the background observables with the
quantization prescription \eqref{ClassQsym}, in which
ordering ambiguities are taken care of by fixing the
parameters of Eq.~\eqref{ClassQsym}. On comparing the mode
functions associated with different basic field variables,
we found that their dynamical equations given in
Eq.~\eqref{d2vs} are inequivalent because of the ambiguous
gravitational potential \eqref{Vs}. Had the background
observables been classical, there would have been no
dynamical ambiguity. Indeed, the dynamics of the mode
functions in classical backgrounds is unique, and the
redefinition of the basic field variables merely transforms
the vacuum state of the perturbation field, as discussed,
e.g., in Ref.~\cite{Grain:2019vnq}. We conclude that it is
the quantum spacetime that is responsible for the dynamical
ambiguity whose physical consequence we study in the
following sections.

In more explicit details, as shown both diagrammatically
through Fig.~\ref{QclCT} and analytically in the Appendix,
the theory described by Eq.~\eqref{HAMpert} is canonically
transformed, classically speaking, into Eq.~\eqref{H2k} and
subsequently quantized to yield the semi-classical
Hamiltonian \eqref{MSqhHsem}. Instead, starting from the
quantum version of Eq.~\eqref{HAMpert} upon substituting
$q\to \hat{Q}$ and $p\to \hat{P}$ at the background level,
but also $\phi_{\bm{k}} \to \hat{\phi}_{\bm{k}}$ and
$\pi_{\phi,\bm{k}} \to \hat{\pi}_{\phi,\bm{k}}$ for the
perturbations, the quantum canonical unitary
transformation~\cite{Peter:2006hx}
\begin{equation}
U_{\bm{k}} = \exp \left( i\alpha  \hat{D}_{v,\bm{k}} \right)
\exp \left( i\beta \hat{v}_{\bm{k}}^2 \right)
\label{Qtrans}
\end{equation}
with $\alpha = \alpha ( \hat{Q},\hat{P} )$ and $\beta =
\beta( \hat{Q},\hat{P} )$ hermitian operators
($\alpha^\dagger=\alpha$, $\beta^\dagger = \beta$) depending
only on the background variables and
\begin{equation}
\hat{D}_{v,\bm{k}} := \frac12 \left( \hat{v}_{\bm{k}} 
\hat{\pi}_{v,\bm{k}} + \hat{\pi}_{v,\bm{k}}
\hat{v}_{\bm{k}} \right),
\label{dilation}
\end{equation}
yields, to second order in perturbations, the quantum
Hamiltonian \eqref{Htempt}. The Appendix includes some
details on the application of the above transformation to
the quantized Hamiltonian \eqref{HAMpert}. The perturbation
variables, to first order, transform as
\begin{align}\begin{split}\label{obst}
\hat{\phi}_{\bm{k}}&=U^\dagger \hat{v}_{\bm{k}} U =
\ex^{-\alpha} \hat{v}_{\bm{k}} + \dots,\\
\hat{\pi}_{\phi,\bm{k}}&=U^\dagger \hat{\pi}_{v,\bm{k}} U=
\ex^{\alpha} \hat{\pi}_{v,\bm{k}} + \left( \beta \ex^\alpha
+ \ex^{\alpha} \beta
\right) \hat{v}_{\bm{k}} +\dots.\end{split}
\end{align}
Setting $\alpha = \ln [\sqrt{1+w} (\hat{Q}/\gamma)^s]$ and
$\beta = -\frac14 \partial_\tau \left[\exp
\left(-2\alpha\right)\right]$ shows that, in the classical
limit, the relations \eqref{obst} indeed become identical
with the relations \eqref{ntoms} [where $Z$ is given by
Eq.~\eqref{Zs}]. If the only quantized variables were the
perturbation variables for which both the classical and the
quantum canonical transformations are linear, then the
quantum motion generated by the respective quadratic
Hamiltonians would be unique irrespective of whether the
choice of perturbation variables is made at the classical or
quantum level (as shown in Ref.~\cite{Grain:2019vnq}). This
is so because there exists a homomorphism between the
Poisson algebra of polynomial phase space observables of
degree no greater than 2 and the brackets of corresponding
operators (see, e.g. \cite{Gotay:1998aw} for more details).
On the other hand, it is known that if a quantum unitary
transformation is nonlinear or/and the Hamiltonian is not
quadratic, then the homomorphism breaks down, which results
in the non-commutativity depicted in Fig.~\ref{QclCT}. In
such cases, the quantum motion depends on whether the choice
of basic variables is made at the classical or quantum
level. If we view the quantum canonical transformation
\eqref{obst}, or the Hamiltonian \eqref{H_real}, from the
perspective of the full Hilbert space
$\mathcal{H}_\textsc{b} \otimes\mathcal{H}_\textsc{p}$,
mixing background and perturbation degrees of freedom, we
note that they are higher order in basic variables as they
involve quantum background and perturbation variables.
Hence, we ascribe the source of the ambiguity to the long
known quantization obstructions (the so-called
Groenewold-Van Hove obstructions
\cite{Groenewold:1946kp,VanHove1951})\footnote{In Ref.
\cite{Gotay:1998aw} the quantized phase space is
$\mathbb{R}^{2n}$, whereas in our case it is
$(q,p,{\phi}_{\bm{k}},{\pi}_{\phi,\bm{k}})\in
\mathbb{R}_{+}\times\mathbb{R}^{3}$. We, however, use the
dilation operator $\hat{D}$ as a basic quantum variable that
combined with $\ln\hat{Q}$ satisfy the canonical commutation
rule and bring our case to that of \cite{Gotay:1998aw}.}.

\section{Getting spectral indices}
\label{numerical}

After having summarized the situation, let us now solve the
quantum dynamics of the perturbation modes, which we do
first numerically and then analytically. We investigate the
amplitude of the perturbations as a function of time for
various wavenumbers $k$, and focus on its dependency on the
free parameter of the conformal parametrization $\chi$.

We shall work in the Heisenberg picture of dynamics and
assume the perturbations to be in a fixed vacuum state that
is the ground state of the quantum Hamiltonian
(\ref{HAMpert}) or (\ref{MSqhHsem}) for all modes of
interest in the large contracting universe
($\eta\to-\infty$). It can be shown that in order for the
vacuum state to be the ground state of the quantum
Hamiltonians in the infinite past $\tau,\eta\rightarrow
-\infty$, the mode functions have to satisfy (up to
irrelevant phase), respectively
\begin{equation}
v^\chi_{k}\big|_{\eta_\textrm{ini}} = \frac{1}{\sqrt{2
k_\chi}} \ \ \ \hbox{and} \ \ \ \frac{\dd
v^\chi_{k}}{\dd\eta}\Big|_{\eta_\textrm{ini}}=
i\sqrt{\frac{k_\chi}{2}}, \label{ini}
\end{equation}
where we used the vanishing of the gravitational potential
in the infinite past\footnote{The initial conditions of
Eq.~\eqref{ini}, are, strictly speaking, approximate in the
general situation. In the fluid parameterization for
instance, the time derivative should read
$$
\frac{\dd v^\F_{k}}{\dd\eta} = i \sqrt{\frac{k_\F}{2}} -
\frac{1}{(1+w)\sqrt{2k_\F}} \frac{1}{ q^{-r_\F}} \frac{\dd
}{\dd \eta} \left( q^{-r_\F}\right),
$$
the second term originating from the fact that it is the
field $\phi$ rather than $v$ that is quantized in this case.
As initial conditions are set for $|\tau| \gg 1$, one can
there use the asymptotic behaviors $\dot{q}/q\sim \tau^{-1}$
and $\eta \sim \tau^{r_\F+r_\C}$, (see
Ref.~\cite{Martin:2021dbz}) to write the extra term in the
time derivative of $v_k^\F$ as
$(3w-1)/[(1+w)(1+3w)\sqrt{k_\F} \eta]$: setting initial
conditions sufficiently deep into the contracting phase
($\eta <0$ and $|\eta| \gg 1$) then permits to neglect such
a term for all parameterizations.}. Equipped with the
initial conditions \eqref{ini}, we now proceed to solve the
mode equation \eqref{d2vs} for the two special cases $s=0$
($\chi=\chi_\C$) and $s=r_\F$ ($\chi=\chi_\F$).

\subsection{Numerical integration}

Since the potential $\mathcal{V}_\chi$ is known explicitly
as a function of the internal time $\tau$ as shown in
Eq.~\eqref{Vchi}, it turns out to be technically more
tractable to switch back to $\tau $ to solve
Eq.~\eqref{d2vs}, even though, for the sake of clarity, we
plot the results in terms of the conformal time,
substituting the numerical value for $\tau(\eta)$ in the
solution. Given the relationship \eqref{eta} between both
times, we have $\dd/\dd \eta = Z^{-2}\dd/\dd\tau$ (we
assume, from now on, that $Z=Z_{r_\F}$) and therefore
\begin{equation}
\frac{\dd^2}{\dd\eta^2} = \frac{1}{Z^4}\frac{\dd^2
}{\dd\tau^2}- \frac{1}{Z^6}\frac{\dd Z^2}{\dd\tau}
\frac{\dd}{\dd\tau},
\end{equation}
so that the perturbation equation of motion for the
Mukhanov-Sasaki variable, namely
\begin{equation}
\frac{\dd^2 v^\chi_k}{\dd\eta^2} + \left[ k^2
-\mathcal{V}_\chi(\eta) \right] v^\chi_k = 0, \label{MSgen}
\end{equation}
reads, plugging the semiquantum solution \eqref{solsem} for
the generic $\chi-$parametrization (with $\chi=\F$ or $\C$)
\begin{widetext}
\begin{equation}
\frac{\dd^2 {v}_{k}^\chi }{\dd x^2}- \frac{2r_\F x}{1+x^2}
\frac{\dd {v}_{k}^\chi}{\dd x} +\left[
\left(\frac{q_\textsc{b}}{\gamma} \sqrt{1 +
x^2}\right)^{4r_\F}(1+w)^2 \tilde{k}_\chi ^2
-\frac{2\omega^2(1-3w)}{9(1-w)^2}\frac{\chi+
x^2}{\left(1+x^2\right)^2}\right] v_{k}^\chi  =0,
\label{ddotv}
\end{equation}
where $k_\chi$ denote the wavenumber in the respective
parametrizations, and we set $\tilde{k}_\chi := k_\chi
/\omega$ as well as $x:=\omega\tau$. In what follows, we
drop the index $\chi$ both on the mode and the wavenumber as
there is no risk of confusion; we will merely specify when
we explicitly calculate for the fluid or conformal
parametrization.
\end{widetext}

Once the initial conditions \eqref{ini} are similarly
expressed in terms of the fluid time $\tau$, the numerical
integration of the above equations allows to follow the
dynamics of the stochastic average
\begin{equation}
\langle \mathcal{R}_{\bm{k}} \mathcal{R}^*_{\bm{k}'} \rangle
= \mathcal{P}_{\mathcal{R}}(k) \delta_{{\bm{k}},{\bm{k}'}},
\label{vRPv}
\end{equation}
providing the amplitude of curvature perturbations (we
follow the convention of Ref.~\cite{Mukhanov:2007zz}, up to
an irrelevant normalisation factor)
\begin{equation}
\delta [k,\tau(\eta)] = \sqrt{\mathcal{P}_{\mathcal{R}}(k)}
= \frac{|v_{k}|}{a}k^{3/2}
= k^{3/2}  |v_k|
\left( \frac{q}{\gamma} \right)^{-r_\C},
\label{prim_amplitude}
\end{equation}
where we made use of the definition \eqref{defGamma} of the
scale factor $a$ in terms of the variable $q$. We focus in
this section to the special cases $\chi = \chi_\F$ and $\chi
= \chi_\C$.

\begin{figure}[t]
\centering
\includegraphics[scale=0.365]{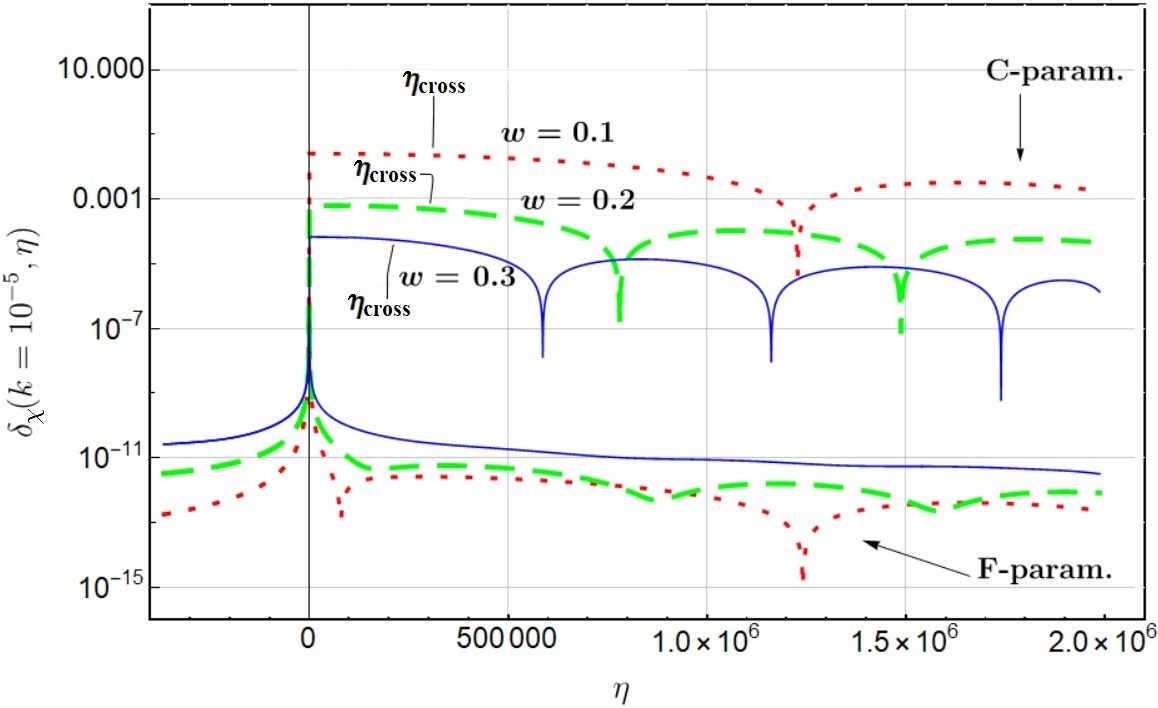}

\caption{Conformal time development of the perturbation
amplitude \eqref{prim_amplitude} for the mode $k=10^{-5}$
and three different equation of state parameters, namely
$w=0.1$ (dotted), $w=0.2$ (dashed), and $w=0.3$ (solid).
Both parametrizations, fluid (bottom, $\chi=\chi_\F$) and
conformal (top, $\chi=\chi_\C$), are shown. Also indicated
is the time $\eta_\text{cross}$ at which the mode exits the
potential, i.e. for which $\mathcal{V} (\eta_\text{cross}) =
k^2$. Here and in the following figures, the background
parameters used are fixed by setting $\kappa_0\to 1$,
$H_\text{sem} = 2^{1+w}$ and $\mathfrak{K}=100$ in
\eqref{solsem}.}

\label{fig:amplitude_ev}
\end{figure}

Although the amplitude is dynamical, it reaches a plateau
right after the bounce when the perturbations have been
amplified, and thus remains roughly constant for a
significant fraction of its period; this corresponds to the
constant (or growing) mode when the perturbation is
dominated by the potential. This is illustrated in
Fig.~\ref{fig:amplitude_ev}, where the dynamics of the
amplitude of a selected mode in both parametrizations and
for three different equation of state parameter $w$. This
constant value of the amplitude right after the bounce is
called the primordial amplitude, and we shall study its
dependence on the wavenumber $k$.

\begin{figure}[t]
\centering \includegraphics[scale=0.235]{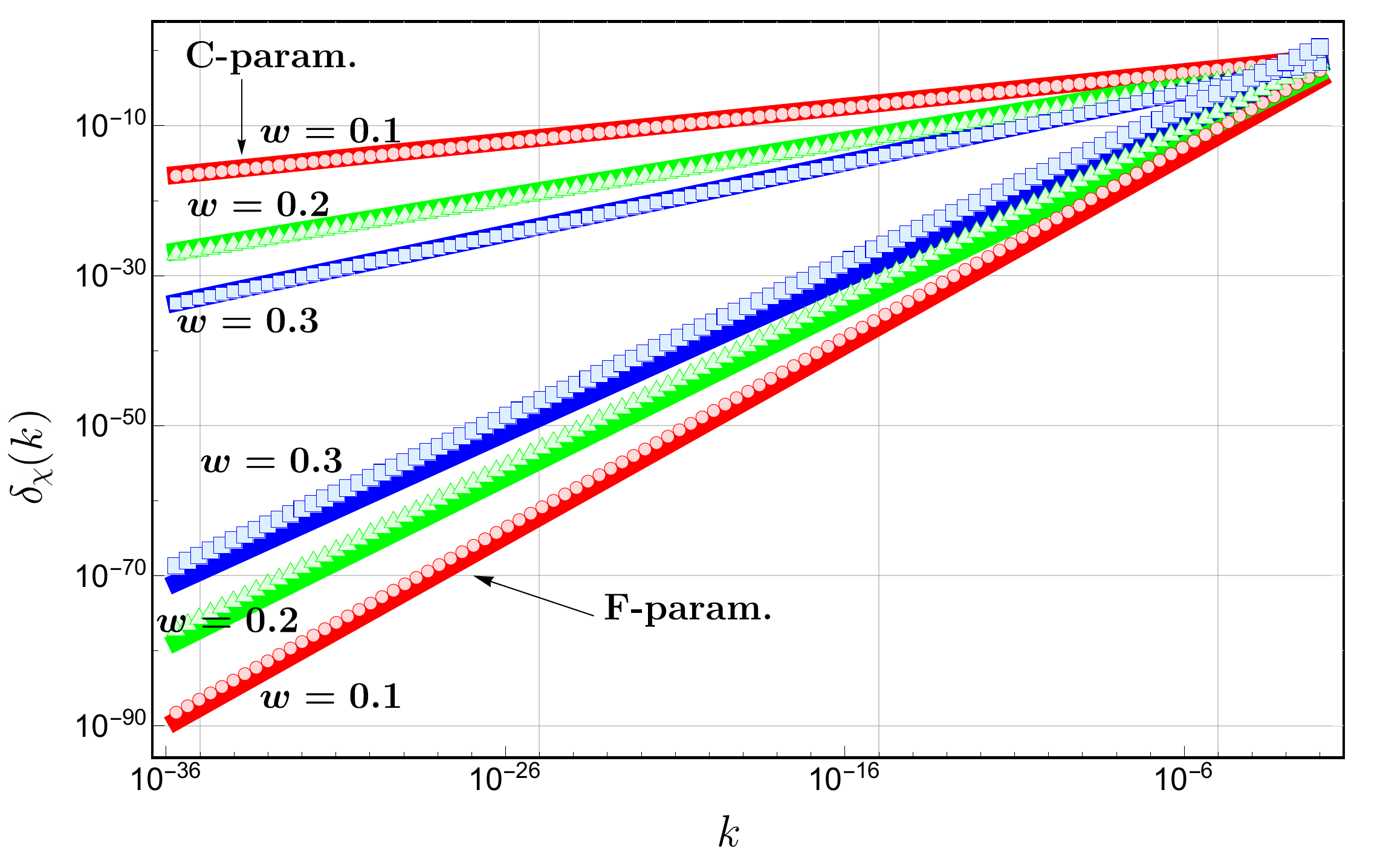}

\caption{Primordial density fluctuation power spectrum
$\delta [k,\tau(\eta_\text{cross})]$ measured at the
crossing time $\eta_\text{cross}$ defined in
Eq.~\eqref{etaC} and shown in Fig.~\ref{fig:amplitude_ev}.
Both the fluid $\F-$(bottom) and the conformal $\C-$(top)
parametrizations are displayed for three different fluids as
in Fig.~\ref{fig:amplitude_ev}, namely $w=0.1$ (circles),
$w=0.2$ (triangles) and $w=0.3$ (squares). The approximate
analytical solutions [Eqs.~\eqref{analytic_spectrum_v} and
\eqref{analytic_spectrum_Phi} below] are shown as
superimposed solid lines for each numerical calculation,
exemplifying the validity of the approximation.}

\label{fig:spectrum}
\end{figure}

We have solved the perturbation equations for many values of
$k$, and calculated their primordial amplitude at the time
of potential-crossing (also  called `exit' time)
$\eta_\textsc{c}$ at which the mode exits the potential,
i.e. for which $\mathcal{V} (\eta_\textsc{c}) = k^2$. The
mode evolutions shown in Fig.~\ref{fig:amplitude_ev} yield
the full spectrum, plotted in Fig.~\ref{fig:spectrum}. One
finds two different power laws for the two different
parametrizations, as expected, thereby emphasizing the
ambiguity at the prediction level. Our analytic estimates
below for the spectral indices represent very accurate fits
for the numerics.

\subsection{Analytical Integration}

We now follow the calculation made in
Ref.~\cite{Peter:2006id} for the case of tensor
perturbations, transcribed to the scalar modes, consisting
in setting a piecewise approximation to the solution of our
general differential equation
\begin{equation}
\frac{\dd^2 v_k}{\dd \eta^2} + \left[ k^2 -
\frac{\left(q^r\right)''}{q^r}\right] v_k =0,
\label{vChi}
\end{equation}
where we set $\alpha^2\to 1$ as we are only interested in
the solution for $r=r_\F$ and $r=r_\C$. We begin by noticing
that long before the bounce, at times for which the modes
are free, i.e. when $k^2\gg |\mathcal{V}|$, the potential is
well approximated by its classical counterpart \eqref{Vcl}.
In this regime, the modes are then given by
\begin{equation}
v_k (\eta) = \sqrt{-\eta} \left[ c_1 H_{\nu}^{(1)}(-k \eta)+ c_2
H_{\nu}^{(2)}(-k\eta) \right],
\label{vHankel}
\end{equation}
where $\nu = \frac{3(1-w)}{2(1+3w)}$ and $H_\nu^{(1,2)}$ are
the Hankel functions of the first and second kinds; the
minus signs appearing in Eq.~\eqref{vHankel} account for the
fact that $\eta < 0$ in the contracting phase. Since, for
$(-k\eta) \gg 1$, one has the asymptotic relations
$H_\nu^{(1)} (-k\eta) \sim \sqrt{\frac{-2}{k\eta\pi}}
\ex^{-i \left[k\eta +\left(\nu + \frac12\right) \right]}$
and $H_\nu^{(2)} (-k\eta) \sim \sqrt{\frac{-2}{k\eta\pi}}
\ex^{i \left[k\eta +\left(\nu + \frac12\right) \right]}$,
the initial conditions \eqref{ini} that impose the
Bunch-Davies vacuum yield $c_1=0$ and $c_2 = \sqrt{\pi/2}\
e^{-i\frac{\pi}{2}(\nu+\frac{1}{2})}$. This implies that at
the time $\eta_\text{in} = -\eta_\text{cross}$ of the
potential crossing $k^2=\mathcal{V}(\eta_\text{cross})$,
namely
\begin{equation}
\eta_\text{cross} = \frac{\sqrt{2(1-3w)}}{(1+3w) k}=:\frac{x_w}{k},
\label{etaC}
\end{equation}
thereby defining the dimensionless variable $x_w$ depending
only on the equation of state $w$, the initial conditions
for the following potential domination era read
\begin{equation}
v_k (\eta_\text{in}) = \frac{C}{\sqrt{k}} \qquad \hbox{and}
\qquad v_k'(\eta_\text{in}) = D \sqrt{k},
\label{matching_2}
\end{equation}
where
\begin{equation}
C = c_2 \sqrt{x_w} H^{(2)}_\nu \left( x_w\right)
\label{C}
\end{equation}
and
\begin{equation}
D = \frac{c_2}{2} \left\{ \frac{H_\nu^{(2)}(x_w)}{\sqrt{x_w}}
+\sqrt{x_w}
\left[
H^{(2)}_{\nu-1}(x_w) - H^{(2)}_{\nu+1}(x_w)
\right]
\right\}.
\label{D}
\end{equation}

{}From the potential crossing conformal time
$\eta_\text{cross}$, one can derive the fluid time
$\tau_\text{cross}$, which depends on the wavenumber $k$ of
a given mode. In the classical approximation, i.e. assuming
this crossing takes place in a regime for which the
potential is well approximated by the classical potential
\eqref{Vcl}, one finds
\begin{equation} \label{initial_tc}
x_\text{cross} = \omega \tau_\text{cross} = \left(
\frac{q_\textsc{b}}{\gamma} \right)^{\frac{2(1-3w)}{1+3w}}
\left[\frac{k}{\omega f(w)}\right]^{-\frac{3(1-w)}{1+3w}},
\end{equation}
where $f(w)=\sqrt{2(1-3w)}/[3(1-w^2)]$.

Once a given mode crosses the potential, we assume the
latter to instantaneously take over the dynamics of the
perturbations, so that Eq.~\eqref{vChi} becomes
(zeroth-order in $k$):
\begin{equation}
\frac{\text{d}^2v_k}{\text{d}\eta^2} -
\frac{\left(q^r\right)''}{q^r}v_k=0,
\label{reduced_equation1}
\end{equation}
whose general solution is found to be
\begin{equation}
v_k = 
\left[ q(\eta) \right]^r
\left\{
A^\chi + B^\chi \int^{\eta} \dd\tilde\eta 
\left[
q(\tilde\eta)
\right]^{-2r}
\right\},
\label{vGenSol}
\end{equation}
where $A^\chi$ and $B^\chi$ are integration constants, later
to depend on $k$ because of the matching with initial
conditions.

In order to use the solution \eqref{vGenSol}, one needs to
express the background motion of $q$ given by
Eq.~\eqref{solsem} as a function of the conformal time
$\eta$. It turns out that Eq.~\eqref{eta} can be integrated
to yield
$$
\eta = (1+w)
\left(\frac{q_\textsc{b}}{\gamma}\right)^{2r_\F} \tau
\mathcal{F}\left[\frac12,-r_\F;\frac32;-(\omega\tau)^2
\right],
$$
with $\mathcal{F}$ an hypergeometric function (see
Ref.~\cite{Martin:2021dbz} for details). We can perform the
integrals in fluid time using the relation \eqref{eta} and
obtain the solutions in terms of $\tau$, absorbing the
choice of the initial time $\eta_0$ into the constants
\begin{equation}
v_k = \left( \frac{q}{\gamma} \right)^r \left\{ A^\chi_k +
\omega \tau
\mathcal{F}\left[\frac12,r-r_\F;\frac32;-(\omega\tau)^2
\right] B^\chi_k \right\}, \label{GenSolr}
\end{equation}
where
$$
A^\chi_k=A^\chi \qquad \hbox{ and } \qquad B^\chi_k =
\frac{B^\chi}{\omega (1+w)}
\left(\frac{q_\textsc{b}}{\gamma}\right)^{2(r-r_\F)}
$$
are unknown functions of the wavenumber $k$.

The solution \eqref{GenSolr} is valid for both the above
parametrizations and, setting $r\to r_\F$ (fluid) or $r\to
r_\C = 1+r_\F$ (conformal) yields
\begin{equation}
v_k^\F = \left(\frac{q}{\gamma}\right)^{r_\F} \left( A_k^\F
+ B_k^\F \tau \right) \label{analytic_solF}
\end{equation}
and
\begin{equation}
v_k^\C = \left(\frac{q}{\gamma}\right)^{r_\C} \left[A_k^\C +
B_k^\C \arctan (\omega\tau) \right]. \label{analytic_solC}
\end{equation}
Given the form \eqref{prim_amplitude} of the amplitude of
curvature perturbations, one needs to evaluate
$(q/\gamma)^{-r_\C} |v_k|$ in the large time limit, which
yields
\begin{equation}
\begin{split}
 \delta  \propto & \left[ A^\chi_k \pm\frac{\sqrt{\pi}}{2}
 \frac{\Gamma\left( r-r_\F-\frac12 \right) }{ \Gamma\left(
 r-r_\F \right)} B^\chi_k\right] \left|\omega\tau
 \right|^{r-r_\C}\\ &+B^\chi_k
 \frac{\ex^{-2i\pi(r-r_\F)}}{1-2(r-r_\F)} \left| \omega\tau
 \right|^{r_\F-r},
\end{split}
\label{deltaFar}
\end{equation}
in which the $\pm$ sign in the first line corresponds to the
sign of $\tau$. One notes that for the values of interest
$r=r_\F$ and $r=r_\C$, this amplitude reads (recall
$r_\C-r_\F=1$)
\begin{equation}
\delta_\F =  \delta  (r_\F) \propto
\frac{A^\F_k}{|\omega\tau|}+B^\F_k, \label{deltarf}
\end{equation}
and
\begin{equation}
\delta_\C =  \delta  (r_\C) \propto \left( A^\C_k \pm
\frac{\pi}{2} B^\C_k \right) - \frac{B_k^\C}{|\omega\tau|}.
\label{deltarC}
\end{equation}
These two distinct solutions both exhibit a constant mode
and a decaying one, and therefore provide a constant
amplitude for the primordial spectrum.

As we assume $w<1/3$, the potential \eqref{Vchi} is positive
definite for $\chi >0$ with a maximum at $\eta=0$. For $\chi
< 0$ on the other hand, this potential has two positive
maxima and a negative minimum at $\eta=0$, so that the modes
cross the potential at four different times (see
Fig.~\ref{FigPots}). Given the symmetry of $V_\chi(\eta)$,
the relevant modes enter the potential for the first time at
$\eta_\text{in} = -\eta_\textsc{c}$ and exit for the last
time at $\eta_\text{out} = \eta_\textsc{c}$ in regions where
it behaves classically (i.e. for $|\omega\tau|\ll 1$).
Fig.~\ref{FigPots} also shows that for a range of values of
$r$, including in particular the fluid parametrization, the
modes also exit and re-enter the potential another time
between those points, when quantum corrections cannot be
neglected. Due to the shape of the potential, there exist
short periods of time during which the value of $k$
dominates over the value of $V_\chi(\eta)$, i.e. in the
neighborhoods of $\pm\eta_0$ defined by
$V_\chi(\pm\eta_0)=0$. It turns out the potential is rather
steep close to those points, with a high negative slope when
the modes exit and a high positive slope when they re-enter
later. For the large wavelengths relevant to the
cosmological framework, this time interval is sufficiently
small that the approximation, assumed in what follows, of
neglecting it altogether, holds. It should be mentioned
however that in that case, the potential becoming negative,
the behaviors of the modes may be quite different; we shall
see below how this should be taken care of.

Combining the matching conditions \eqref{matching_2} at the
time \eqref{initial_tc} with the potential-domination
solutions \eqref{analytic_solF} and \eqref{analytic_solC}
yields the coefficients $A^\F$, $A^\C$, $B^\F$, and $B^\C$
as functions of $k$. To leading order in $k\ll 1$, one gets
\begin{align}
A^\F_k =\left(
\frac{q_\textsc{b}\omega f}{\gamma} 
\right)^{\frac{1-3w}{1+3w}}
\left[r_\C C + (1+w)f D
\right] k^{-\frac{3(1-w)}{2(1+3w)}}
\label{AF}
\end{align}
and
\begin{align}
B^\F_k =f \left(
\frac{q_\textsc{b}\omega  f}{\gamma}
\right)^{-\frac{1-3w}{1+3w}}
\left[r_\F C + (1+w)f D
\right]
k^{\frac{3(1-w)}{2(1+3w)}}
\label{BF}
\end{align}
We note that $A^\F_k$ and $B^\F_k$ have inverse $k$
dependence, so that, in the large wavelength limit, $A^\F_k
\gg B^\F_k$. For the conformal case, one finds
\begin{align}
A_k^\C = \frac{\pi}{2} B_k^\C + C \left(
\frac{q_\textsc{b}}{\gamma} \right)^{-\frac{2}{1+3w}}
k^{\frac{3(1-w)}{2(1+3w)}} \label{ACBC}
\end{align}
and $B_k^\C = \gamma A_k^\F /q_\textsc{b}$, relations that
can also be obtained by merely equating
\eqref{analytic_solF} and \eqref{analytic_solC} and their
derivatives at the matching point; we kept the subdominant
term in $B_k^\C $ in \eqref{ACBC} for further convenience.

\begin{figure}[t]
\centering
\includegraphics[scale=0.25]{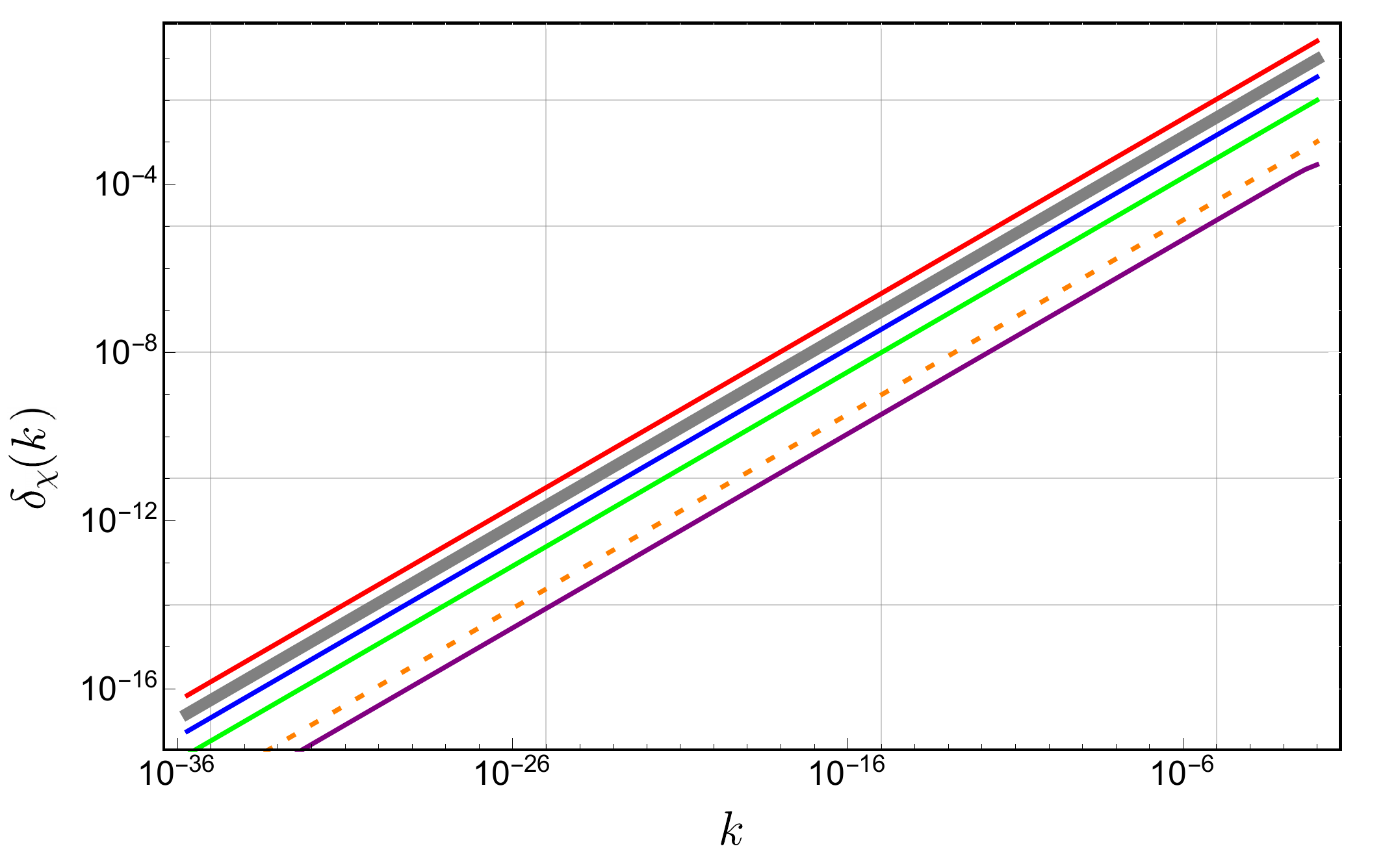}

\caption{Numerical primordial amplitude power spectrum at
the crossing point for $w=0.1$ and various values of $\chi$;
in that case, $\chi_\F=-1.35$ and $\chi_\C \simeq 3.86$.
From top to bottom: $\chi=9$, $\chi=\chi_\C$ (thick line),
$\chi=2$, $\chi=-0.1$, $\chi=-1.5$ (dashed), and
$\chi=-1.3$. This illustrates the fact that although the
amplitude depends on $\chi$, the index remains given by
\eqref{nC} provided $\chi\not=\chi_\F$ (not shown). For
$\chi > \chi_\F$, the amplitude is seen to decrease as
$\chi\to\chi_\F$. It increases again for $\chi < \chi_\F$
(the dashed line).}

\label{fig:spectrum_chi} 
\end{figure}

The last step consists in substituting the above expressions
into the primordial amplitude spectrum
\eqref{prim_amplitude}, keeping the highest order terms in
$k\ll 1$, assuming $0\leq w \leq \frac13$. This yields
\begin{equation}
\delta_\chi (k) = A(w,\chi) \omega f
\left(\frac{k}{\omega f}\right)^{n(w,\chi)},
\label{deltachi}
\end{equation}
where the amplitudes
\begin{align}
A(w,\chi_\F) = \left( \frac{q_\textsc{b}}{\gamma}
\right)^{-\frac{2}{1+3w}} \left|\left( r_\F+r_\C\right) C +
2 (1+w) f D\right| \label{analytic_spectrum_Phi}
\end{align}
and
\begin{align}
A(w,\chi_\C) = \pi \left(
\frac{q_\textsc{b}}{\gamma}\right)^{-\frac{6w}{1+3w}}
\left|r_\C C + (1+w) f D \right|
\label{analytic_spectrum_v}
\end{align}
as well as the spectral indices
\begin{align}
n(w,\chi_\F) = \frac{3(1+w)}{1+3w}
\label{nF}
\end{align}
and
\begin{align}
n(w,\chi_\C) = \frac{6w}{1+3w}
\label{nC}
\end{align}
differ to yield effectively distinguishable predictions: the
power spectrum being the square of the fractional energy
density, i.e. $\mathcal{P}_\textsc{s} (k) = \delta_\chi^2
\propto k^{n_\textsc{s} -1}$, one finds two different power
indices, namely that given by the fluid parametrization
$n^\F_\textsc{s} - 1 = 6 (1+w_\F)/(1+3w_\F)$ (i.e. $w_\F
\sim -0.988$ to agree with the CMB Planck
data~\cite{Planck:2018jri}), and $n^\C_\textsc{s} -1= 12
w_\C/(1+3w_\C)$ for the conformal one (i.e. $w_\C\sim
-2.9\times 10^{-3}$); it is the latter expression which is
usually assumed~\cite{Peter:2006hx}. Fig.~\ref{fig:spectrum}
shows, for various values of the equation of state parameter
$w$, by superimposing the results, that the predicted
spectra \eqref{analytic_spectrum_Phi} and
\eqref{analytic_spectrum_v}, agree with the numerical
calculation.

\section{A tale of two indices}
\label{tale}

Solving Eq.~\eqref{MSgen} for arbitrary values of $\chi$ is
not possible because of the parameter $\alpha_\chi$ in
Eq.~\eqref{qrdd}, as it is only for $\alpha_\chi=1$, i.e.
for $\chi \in \{\chi_\F,\chi_\C\}$, that the analytic
solutions \eqref{vGenSol} are valid. Given that this
parameter comes from the quantization process and is thus
seemingly arbitrary, one may reasonably worry that the
spectral index of scalar perturbations might depend on its
exact value, the theory therefore loosing its predictive
power. Indeed, for the two values for which one can solve
the mode equation, one already obtains an ambiguity as the
two spectral indices \eqref{nF} and \eqref{nC} are both
possible predictions.

\begin{figure}[h]
\centering
\includegraphics[scale=0.305]{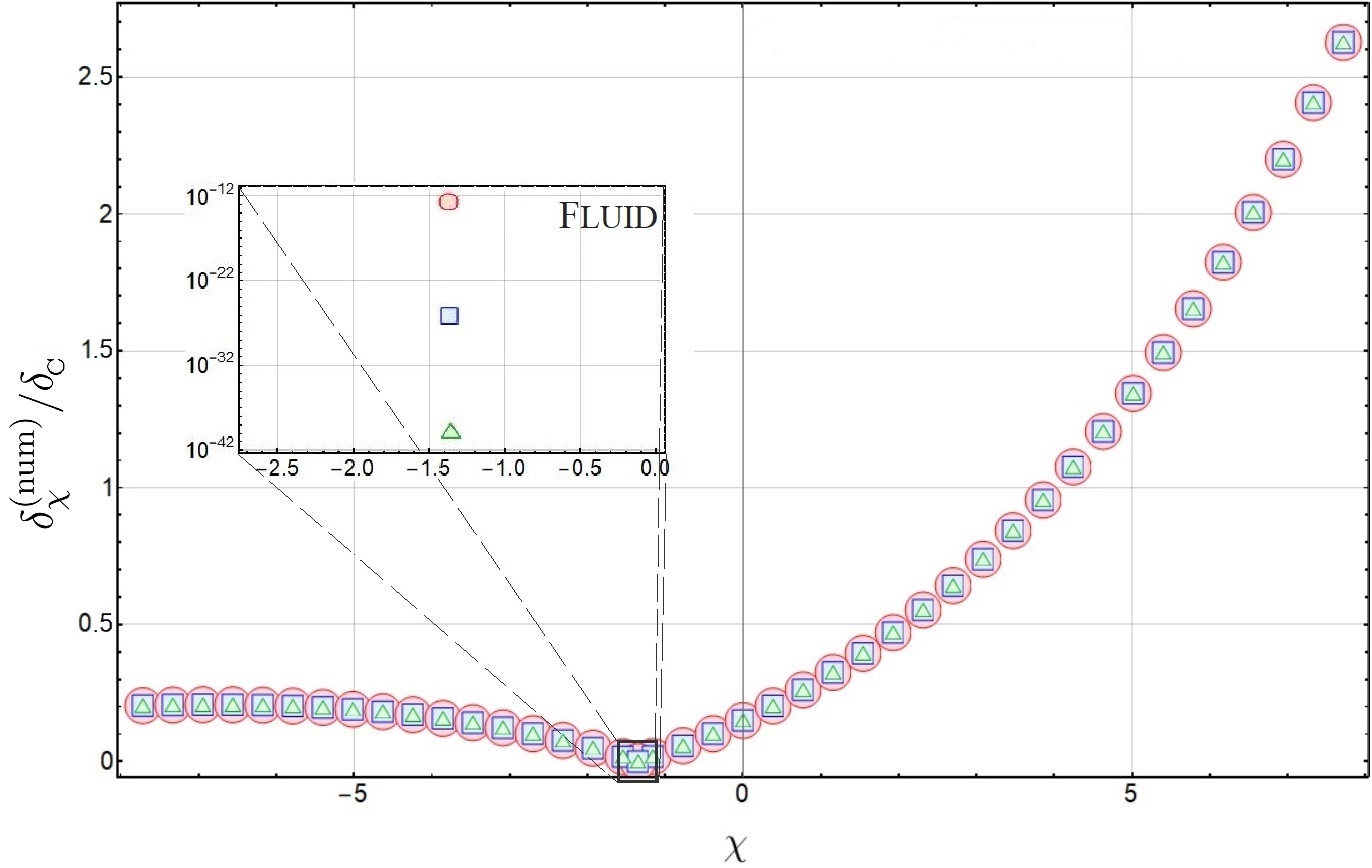}

\caption{Ratio $ \delta_\chi^{\text{(num)}}/ \delta_\C$
between the numerical primordial amplitude power spectrum 
with its analytic counterpart of the conformal
parametrization given by Eq.~\eqref{deltachi} as a function
of $\chi$ for $w=0.1$ and three different modes, namely
$k_{\C}=10^{-15}$ (circles), $k_\C=10^{-25}$ (squares) and
$k=10^{-35}$ (triangles). The coincidence of the three
curves for all values of $\chi$ expect $\chi_\F$ indicates
that the spectral index is indeed $n_\textsc{s}^\C$ given by
\eqref{nC} and independent of $\chi$. At
$\chi=\chi_\F=-1.35$, the zoom shows three distinct points,
exhibiting that the spectral index differs, being then given
by \eqref{nF}.}

\label{fig:ratioVsChiPlusZoom} 
\end{figure}

\subsection{Numerical facts}

Fig.~\ref{fig:spectrum_chi} shows the fluctuation $ \delta $
as a function of the wavenumber $k$ for various values of
$\chi$. One immediately notices that although the amplitude
depends on $\chi$, decreasing with $\chi$ until $\chi_\F$
and then increasing again for $\chi<\chi_\F$, we obtain the
same power law for the spectral index that in the special
case $\chi=\chi_\C$ \eqref{nC} for all the possible values
that $\chi$ can take within the conformal parametrization in
its generic form \eqref{VchiK}. In order to clarify this
point, we plot, in Fig.~\ref{fig:ratioVsChiPlusZoom}, the
ratio between the numerical spectrum obtained for whatever
value of $\chi$ and that provided by our analytical
approximation \eqref{deltachi} for $\chi=\chi_\C$. This plot
is but an example for a given value of $w$, we recovered the
same generic image for all values we investigated.

\begin{figure}[h]
\centering
\includegraphics[scale=0.24]{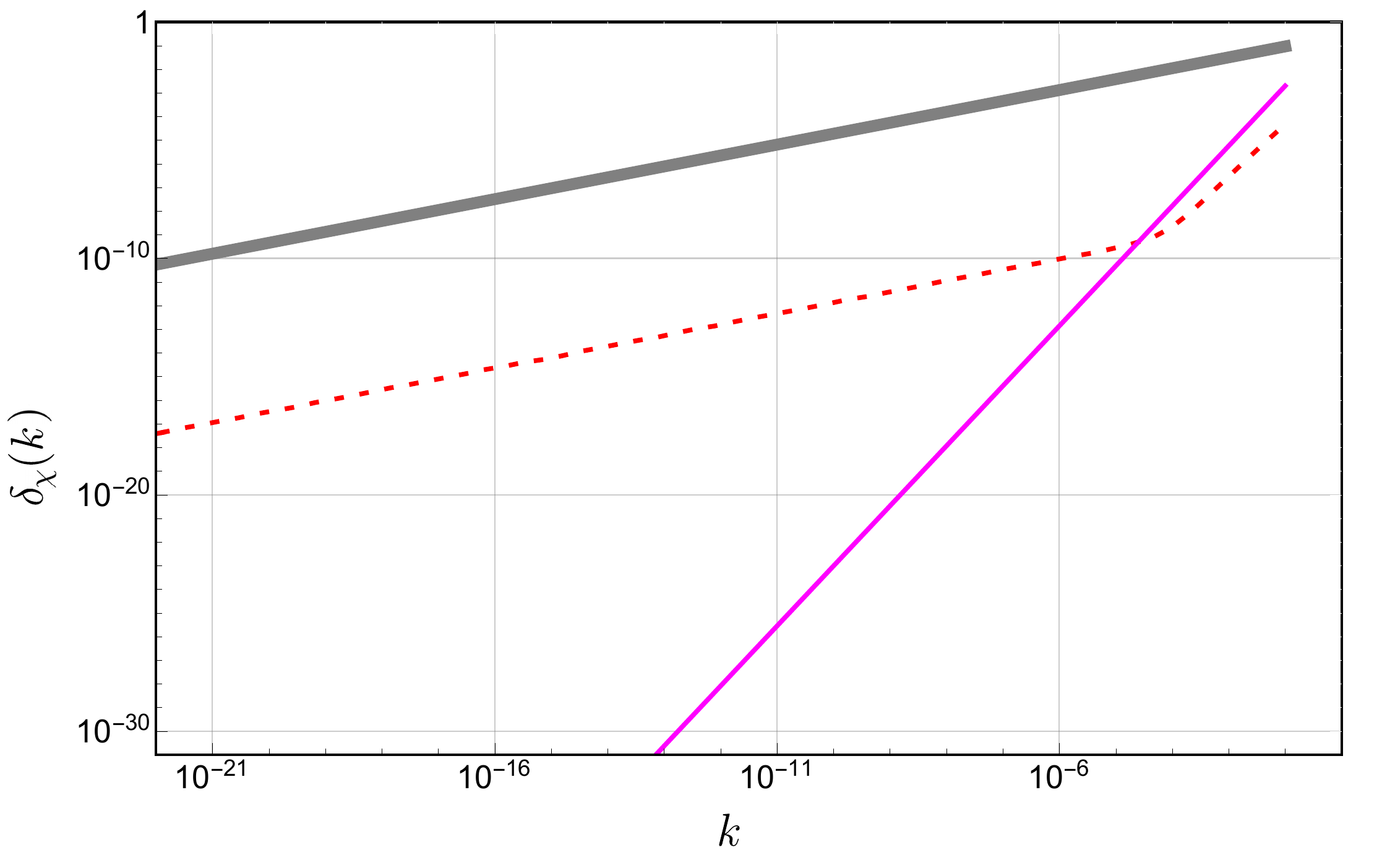}

\caption{Primordial density fluctuation spectrum for
$w=0.1$: the conformal case \eqref{analytic_spectrum_v}
$\chi=\chi_\C=3.86$ is shown as the thick line above, and
the fluid case $\chi=\chi_\F$ as the thin bottom line. The
dashed line represents a case close to the Fluid case with
$\chi=\chi_\F-10^{-6}$.}

\label{fig:closetoFluid} 
\end{figure}

What is also seen in Fig.~\ref{fig:ratioVsChiPlusZoom} is a
generalisation of Fig.~\ref{fig:spectrum_chi}, namely that
the spectral index of scalar perturbations is generically
given by \eqref{nC} expect in the case of $\chi=\chi_\F$.
The fluid parametrization thus corresponds to the minimum
amplitude possible and a different scalar index. In fact, we
find that it is understandable as the situation in which the
conformal amplitude merely vanishes, leading to the
subdominant fluid amplitude being the only one, thereby
dominating the full spectrum.

We tested this hypothesis by calculating the spectrum as a
function of the wavenumber for various values of $\chi$
close to the fluid case $\chi_\F$. The result is illustrated
in Fig.~\ref{fig:closetoFluid} in which the power spectrum
is calculated numerically for a value of the parameter
$\chi$ very close to $\chi_\F$, superimposed with the
analytic solutions for the conformal and the fluid cases. It
can be seen that the full spectrum somehow interpolates
between both cases, following the fluid power law for large
wavenumbers and the conformal power law for smaller
wavenumbers. This suggests that the full spectrum contains
both power law terms, the amplitude depending on
$\chi-\chi_\F$ \JM{for \eqref{nC}}: as both power laws are
positive, when $k$ decreases, the contribution due to
$n_\textsc{s}^\F$ becomes smaller compared to that due to
$n_\textsc{s}^\C$, so the latter finally dominates entirely
for very small wavenumbers.

\subsection{A sharp transition}\label{sharp}

A better analytical understanding of the numerical results
of the previous section can be achieved by investigating
more closely the potential \eqref{Vchi} seen as a function
of time and $\chi$.

Let us first assume that the reference value for $\chi$ is
given by the conformal one $\chi_\C$. The potential for any
value of $\chi$ can be written as $V_\chi = V_\C + \delta
V$, where
\begin{equation}
\delta V = \frac{2\omega^2}{9Z^4} \frac{1-3w}{(1-w)^2}
\delta\chi\left[1+\left(\omega \tau\right)^2\right]^{-2}
\label{deltaV1}
\end{equation}
and $\delta\chi= \chi-\chi_\C$. Plugging the definition of
$Z$ and looking at the large time limit of the full
potential, one finds that $\lim_{\omega\tau\to\infty} \delta
V \ll \lim_{\omega\tau\to\infty} V_\chi$, so that the main
contribution of $\delta V$ is around the bounce time, namely
around $\tau =0$. As a function of $\tau$, one indeed finds
\begin{equation}
\delta V = \frac{2\omega^2}{9 (1+w)^2}
\left(\frac{q_\textsc{b}}{\gamma}\right)^{-4r_\F}
\frac{1-3w}{(1-w)^2} \delta\chi\left[1+\left(\omega
\tau\right)^2\right]^{-2 r_\C}. \label{deltaV2}
\end{equation}
Since we focus on the cases $0\leq w \leq \frac13$, one has
$\frac23\leq r_\C \leq 1$, so that, compared to $V_\C$, one
can approximate $\delta V$ as though its contribution is
localized entirely at the bounce, i.e. we replace $\delta V$
by
\begin{equation}
\delta V_\text{approx} = \Upsilon \delta (\eta),
\label{deltaVdelta}
\end{equation}
where we assume the coefficient $\Upsilon$ takes the form
$$
\Upsilon = \varpi \int_{-\infty}^{\infty}\delta
V \dd \eta,
$$
with $\varpi = \varpi(w) \approx 1$ (we shall evaluate it
below), and the integral can be calculated to yield
\begin{equation}
\Upsilon =\frac{2\sqrt{\pi}\omega(1-3w)}{9(1+w)(1-w)^2}
\left(\frac{q_\textsc{b}}{\gamma}\right)^{-2r_\F}
\frac{\Gamma\left(r_\C+\frac12\right)}
{\Gamma\left(r_\C+1\right)} \varpi(w) \delta\chi,
\label{deltaVdelta}
\end{equation}
transforming Eq.~\eqref{reduced_equation1} into
\begin{equation}
\frac{\text{d}^2v_k}{\text{d}\eta^2} -
\left[ \frac{\left(q^{r_\C}\right)''}{q^{r_\C}} +
\Upsilon\delta(\eta)
\right]v_k=0,
\label{d2vdelta}
\end{equation}
whose solution is given by Eq.~\eqref{analytic_solC} on both
sides of the bounce, only with different parameters $A^{\C
<}_k=A^{\C}_k$, $B^{\C <}_k=B^{\C}_k$ for the contracting
phase [given by Eq.~\eqref{ACBC}] and $A^{\C >}_k$, $B^{\C
>}_k$ for the expanding phases.

Assuming continuity of $v_\chi$ at $\eta=0$ yields $A^{\C
>}_k=A^\C_k$, and integrating Eq.~\eqref{d2vdelta} around
the bounce provides the discontinuity in the time derivative
as
\begin{equation}
v'_k(0^+) - v'_k(0^-) = \Upsilon v(0),
\end{equation}
leading to
\begin{equation}
B^{\C >}_k = B^{\C}_k -\sqrt{\pi}
\frac{\Gamma\left(r_\C+\frac12\right)}
{\Gamma\left(r_\C+1\right)}
\frac{\varpi\delta\chi}{\chi_\F - \chi_\C},
\end{equation}
as can be shown by direct evaluation of $\chi_\F - \chi_\C$
as a function of $w$.

Plugging the above value of $B^{\C >}_k$ into the definition
\eqref{prim_amplitude} and using the mode solution
\eqref{analytic_solC}, one finds that the power spectrum now
consists in two contributions, namely $ \delta  = k^{3/2}
\left( D + S \right)$, with the dominant term given by
\begin{equation}
D = \pi | B_k^\C | \left[ 1 - \frac{\pi^{3/2}}{4} 
\frac{\Gamma\left(r_\C+\frac12\right)}
{\Gamma\left(r_\C+1\right)}
\frac{\varpi\delta\chi}{\chi_\F - \chi_\C}\right],
\label{Cdominant}
\end{equation}
while the subdominant term reads
\begin{equation}
S = S_\textsc{n} |C| \left( 
\frac{q_\textsc{b}}{\gamma}
\right)^{-\frac{2}{1+3w}}
k^{\frac{3(1-w)}{2(1+3w)}},
\end{equation}
with normalisation
\begin{equation}
S_\textsc{n} =\left[ 1 - \frac{\pi^{3/2}}{2}
\frac{\Gamma\left(r_\C+\frac12\right)}
{\Gamma\left(r_\C+1\right)}
\frac{\varpi\delta\chi}{\chi_\F - \chi_\C}\right].
\end{equation}
Note that the two modes are obtained only when one keeps the
otherwise negligible contribution in Eq.~\eqref{ACBC}; this
is why we kept it in the first place.

Eq.~\eqref{Cdominant} shows that, for a fixed unique
parameter $\varpi(w)$ of order unity that best approximates
$V_\text{approx}$ to $\delta V$, there is one and only one
value of $\chi$, for which the dominant mode vanishes,
thereby explaining our numerical findings. Therefore, one
can assume that the best fit is given by
\begin{equation}
\varpi(w) = \frac{4}{\pi^{3/2}}
\frac{\Gamma\left(r_\C+1\right)}
{\Gamma\left(r_\C+\frac12\right)},
\end{equation}
leading the dominant term to only vanish for $\chi=\chi_\F$,
in agreement with the above results. For $w$ in our range,
this lies between $\varpi(0) \approx 0.7$ and $\varpi(1/3)
\approx 0.8$, i.e. a number of order unity as expected.

With the power spectrum \eqref{Cdominant} vanishing, there
remains the subdominant piece, which happens to lead to $
\delta  \propto k^{n(w,\chi_\F)}$ [see Eq.~\eqref{nF}]. This
reproduces exactly the features observed in the previous
section and illustrated in Fig.~\ref{fig:closetoFluid},
namely that as $\chi\to\chi_\F$, the dominant amplitude
coefficient becomes smaller and smaller and thus comes to
actually dominate over the subdominant one only for smaller
wavenumbers. In the limit $\chi =\chi_\F$, the coefficient
exactly vanishes and the subdominant piece, then being the
only one, becomes the only relevant spectrum.

\section{Conclusion}
\label{conclusion}

In this work, we performed a detailed examination of the
dynamical ambiguity that naturally arises in models of the
primordial universe, in which both the cosmological
background and the perturbations are quantized. A previous
paper had found that quantizing the background and writing a
semiquantum trajectory approximation leads to two different
potentials for the perturbations, thereby rendering the
theory effectively ambiguous and potentially unpredictive.

The model presented in this work is only academic in that it
describes a universe whose dynamics is driven by a fluid at
all times and fails at reproducing the CMB data: to do so,
it would require either $w\sim -2.9\times 10^{-3}$ for the
conformal case, and $w\sim -0.988$ in the fluid case. Both
being negative, the corresponding models are plagued with
incurable instabilities; the model we have discussed here
assume $0\leq w\leq \frac13$ so as to avoid such
instabilities. It merely serves as an illustration of the
ambiguity and its resolution: one can expect that the same
techniques using a scalar field should lead to similar
results.

At the classical level, one can build an infinite number of
acceptable and equivalent (i.e. related by canonical
transformations) perturbation variables which, upon
quantization, lead to a priori different quantum theories:
obtaining the Mukhanov-Sasaki variable through performing
the canonical transformation either before quantization or
after the semiquantum trajectory is obtained yields
inequivalent potentials and therefore, one would have
guessed, to different predictions for the expected spectrum.
We found the astonishing result that despite the presence of
a continuous parameter describing the various possible
potentials, there are only two possible predictions for the
spectral index.

The conformal parametrization reproduces the usual spectrum
with $n_\textsc{s}^\C$, the amplitude depending on the
parameters of the semiquantum trajectory.

There exists, because of the ambiguity, another possible
spectral index, stemming from using the fluid
parametrization. We found however this prediction to be very
special: within our family of potentials depending on one
parameter $\chi$, we found that all values of $\chi$ predict
the same (conformal) spectral index, except when $\chi\to
\chi_\F$ exactly, in which case one gets a different
spectrum with no parameter and a single well-defined
amplitude.

The fluid case can be explained as leading to the
subdominant contribution in the spectrum, the amplitude of
the dominant term vanishing for the special value $\chi\to
\chi_\F$. In that sense, it represents a set of measure zero
in the general $\chi \in \mathbb{R}$, so that one can either
deduce that the conformal case is the generic, and therefore
the correct one, or, on the contrary, that the fluid case
being so special should represent the correct prediction.

The conformal spectrum depends on the ratio $r$ between the
3D compact manifold and the observable universe, as well as
on the parameter $\mathfrak{K}$ coming from quantization of
the background. There exists some amount of degeneracy
between those. The tensor index should be calculated to
raise this degeneracy. It is to be expected that a similar
behavior will be observed.

In quantum bounce models, the Mukhanov variable belongs to
the class of variables that yields the generic prediction
for primordial amplitude spectrum. In this sense, its use in
semi-classical theories like inflation can be given a deeper
justification. On the other hand, it is not the only one in
that class. Moreover, which is the most convenient choice of
variable it depends on the choice of internal time employed
in quantization. Hence, the Mukhanov variable constitutes a
valid but no longer a preferred choice in quantum bounce
models.

\begin{acknowledgments}

P.M. and J.C.M. acknowledge the support of the National
Science Centre (NCN, Poland) under the research grant
2018/30/E/ST2/00370.

\end{acknowledgments}

\appendix*

\section{Quantum canonical transformation}

Let us start with the Hamiltonian \eqref{HAMpert} and
implement in a quantum way the canonical transformation
taking the fluid variable $\phi_{\bm{k}}$ into its classical
conformal-equivalent $v_{\bm{k}}$. In what follows, we
assume the fields to be real, or more precisely we focus on
the real parts of the fields, the imaginary parts being
treated in an exactly similar way so that the full
Hamiltonian is merely obtained by adding the real and
imaginary components. Besides, for the sake of simplicity,
we drop the wavenumber indices, so that our starting point
\eqref{HAMpert} is
\begin{equation}
H_\phi^{(2)} = \frac{1}{2}\left[ \pi_\phi^2 + w(1+w)^2
\left(\frac{q}{\gamma}\right)^{4 r_\F} k^2 \phi^2 \right],
\label{H_real}
\end{equation}
which is quantized through the replacements
$\phi\to\hat{\phi}$, $\pi_\phi\to\hat{\pi}_\phi$ and
$q\to\hat{Q}$, the background Hamiltonian being taken to be
$\hat{H}^{(0)} = \frac12 \hat{P}^2 + K \hat{Q}^{-2}$; that
is, we set $\kappa_0\to \frac14$ and $\mathfrak{c}_0=2K$ in
this appendix; in any case, these variables do not enter in
the subsequent results and can therefore be given whichever
value to simplify the calculations. As all the variables are
now quantum in what follows, we remove all the hats over the
operators.

Let us now perform the quantum version of the canonical
transformation \eqref{ntoms}. Switching to a different basis
perturbation variable $v$ is achieved by introducing a
unitary transformation
$U[v(\phi,\pi_\phi),\pi_v(\phi,\pi_\phi)]$ of
$\mathcal{H}_\textsc{b} \otimes\mathcal{H}_\textsc{p}$, the
Hilbert space of states mixing background and perturbation
variables. It can be chosen as
\begin{equation}
U = \ex^{i\alpha (Q,P) D_v} \ex^{i\beta (Q,P) v^2},
\label{Utransf}
\end{equation}
with $D_v = \frac12 \left( v \pi_v + \pi_v v\right)$, so
that $\left[ v,D_v \right] = i v$; the two Hermitian
operators $\alpha$ and $\beta$ depend only on the background
variables $Q$ and $P$, and therefore implicitly on time.

Under the transformation \eqref{Utransf}, the field becomes
\begin{equation}
v = U\phi U^\dagger\quad \Longrightarrow \quad \phi =
U^\dagger v U = \ex^{-\alpha} v + \cdots, \label{Phiv}
\end{equation}
where the ellipsis means higher order terms which we neglect
-- in this case $\mathcal{O}(v^3)$; as a general matter,
from that point on, we assume the leading order and do not
write explicitly the order of the missing terms.

Similarly, the dilation operator transforms into
\begin{equation}
D_v = U D_\phi U^\dagger \quad \Longrightarrow \quad D_\phi
= U^\dagger D_v U = D_v +2 \beta v^2 \label{Dv}
\end{equation}
which, given our choice \eqref{Utransf}, happens to be exact
at all orders. One gets the canonical momentum as
$$
\pi_v = \frac12 \left( D_v v^{-1} + v^{-1} D_v \right),
$$
and a similar definition for $\pi_\phi$ in terms of the
equivalent $D_\phi$. This leads to
\begin{equation}
\pi_\phi = \ex^{\alpha} \pi_v +
\left( \beta \ex^\alpha + \ex^{\alpha} \beta 
\right) v,
\label{piphipiv}
\end{equation}
remembering that $[ \alpha,\beta ]\neq 0$. 

The operators $\phi$ and $v$ obey the usual Heisenberg
equation
$$
\frac{\dd \phi}{\dd\tau} = i\left[ H_\phi^{(2)}(\phi,
\pi_\phi), \phi \right] \quad \text{and} \quad \frac{\dd
v}{\dd\tau} = i\left[ H_\textsc{t}^{(2)}(v,\pi_v) , v
\right],
$$
with the Hamiltonian given by \eqref{H_real}, while its
transformed counterpart $H_\textsc{t}^{(2)}(v,\pi_v)$ needs
to be determined.

Using \eqref{Phiv} and expanding the time derivative, one
gets
\begin{equation}
\frac{\dd v}{\dd\tau} = i \left[ 
U H_\phi^{(2)} U^\dagger + i U 
\frac{\dd U^\dagger}{\dd\tau}, v\right],
\label{vdot}
\end{equation}
thereby defining the transformed Hamiltonian
$H_\textsc{t}^{(2)}$. Now, the time derivative of $U$ reads
$$
\frac{\dd U^\dagger}{\dd\tau} = i \left[ H_\phi^{(2)},
U^\dagger \right] + \frac{\partial U^\dagger}{\partial\tau},
$$
so that, from \eqref{vdot}, one gets a cancellation of the
first term to yield the total Hamiltonian
$H_\textsc{t}^{(2)}(v,\pi_v)$ for the perturbations in terms
of the variables $v$ and $\pi_v$, namely
\begin{equation}
H_\textsc{t}^{(2)}(v,\pi_v) = iU(v,\pi_v)
\dot{U}^\dagger(v,\pi_v) + H_\phi^{(2)}(U^\dagger {v}
U,U^\dagger {\pi}_{v} U), \label{effectiveH}
\end{equation}
a dot meaning partial derivative with respect to the time
$\tau$ and the last term being calculable from
\eqref{H_real} through the replacements \eqref{Phiv} and
\eqref{piphipiv}.

Expanding \eqref{Utransf} in powers of $v$ and $D_v$, the
first term in \eqref{effectiveH} reads
\begin{equation}
i U \dot{U}^\dagger = -\dot{\alpha} D_v
-\left(\dot{\beta}+\dot{\alpha}\beta+\beta\dot{\alpha}\right)v^2 + \cdots. \label{Udot}
\end{equation}
\noindent We now set $\alpha = \alpha(Q) = \ln Z_s$, with
$Z_s$ the operator counterpart of \eqref{Zs}. The time
evolution of $\alpha$ is given by the Heisenberg equation
\begin{equation}
\dot{\alpha} = -i \left[\alpha,H^{(0)}\right] = -i \left[
\alpha(Q),\frac12 P^2 +\frac{K}{Q^2} \right],
\end{equation}
which yields 
\begin{equation}
\frac{\dd}{\dd\tau} \underbrace{\ln \left[ \sqrt{1+w} \left(
\frac{Q}{\gamma} \right)^s \right]}_{\alpha(Q)} =
\dot{\alpha} = \frac{s}{2} \left(i Q^{-2} + 2 Q^{-1} P
\right), \label{dotalpha}
\end{equation}
in which we used $\left[ f(Q), P \right] = i\dd f/\dd Q$.
This also yields the commutation relation
\begin{equation}
\left[ \dot{\alpha},\alpha \right] = - i s^2 Q^{-2} = -is^2
\frac{(1+w)^{1/s}}{\gamma^2} \ex^{-2\alpha/s} =g(\alpha)
\label{galpha}
\end{equation}
between $\alpha$ and its time derivative.

In order for the transformation \eqref{Utransf} to be
canonical, the term proportional to $D$ in the transformed
Hamiltonian must disappear. This requires its coefficient to
vanish, a condition which is fulfilled provided $\alpha$ and
$\beta$ satisfy
\begin{equation}
\dot{\alpha} = \ex^\alpha \beta \ex^\alpha + \frac12 \left(
\ex^{2\alpha} \beta + \beta \ex^{2\alpha} \right).
\label{dotalphabeta}
\end{equation}
Plugging Eq.~\eqref{dotalphabeta} into \eqref{galpha}, one
gets
$$
g(\alpha) = \ex^\alpha \left[ \beta, \alpha \right] +\frac12
\ex^{2\alpha} \left[ \beta, \alpha \right] + \frac12 \left[
\beta, \alpha \right] \ex^{2\alpha},
$$
showing that the commutator $\left[ \alpha,\beta \right]$
can only depend on $\alpha$. All terms on the r.h.s of the
above equation commuting, it is easy to solve for the
commutator. One gets
$$
\left[ \beta, \alpha \right] = \frac{1}{2} \ex^{-2\alpha}
g(\alpha) = -\frac{is(1+w)^{1/s}}{2\gamma^2}
\ex^{-2\alpha(1+s)/s},
$$
so that $\beta = \frac{1}{2} \ex^{-2\alpha} \dot{\alpha} +
\beta_0 (\alpha)$. The function $\beta_0$ can be found by
replacing the above $\beta$ into \eqref{dotalphabeta}. This
yields $\beta_0 = -\frac{1}{2} g(\alpha) \ex^{-2\alpha}$,
and finally
\begin{equation}
\beta =\frac{1}{2} \ex^{-2\alpha} \left[ \dot{\alpha} -
g(\alpha) \right] = -\frac{1}{4} \frac{\dd
\ex^{-2\alpha}}{\dd\tau}, \label{betaalpha}
\end{equation}
which is expressed in terms of the background operators as
\begin{equation}
\beta = \frac{s}{4 (1+w)}
\left(\frac{Q}{\gamma}\right)^{-2s} \left[i (1+2s) Q^{-2} +
2 Q^{-1} P \right]. \label{betaQP}
\end{equation}
One finally gets the Hamiltonian $H_\textsc{t}^{(2)}$,
namely
\begin{widetext}
\begin{equation}
H_\textsc{t}^{(2)} = \frac{i}{2} \left[ \beta, \ex^{2\alpha}
\right] + \frac12 \ex^{2\alpha} \pi_v^2 +\frac12 \left[
\left( \beta \ex^\alpha + \ex^\alpha \beta \right)^2
-2(\dot{\alpha}\beta+\beta\dot{\alpha}) - 2\dot\beta +
w(1+w)^2 \left( \frac{Q}{\gamma} \right)^{4r_\F} k^2
\ex^{-2\alpha} \right]v^2. \label{Htempt}
\end{equation}
The first term in Eq.~\eqref{Htempt} depends only on the
background variables. It can in fact be evaluated as
$$
\frac{i}{2} \left[ \beta, \ex^{2\alpha} \right] =
\frac{s^2}{2 Q^2},
$$
which merely implies a correction to the otherwise unknown
parameter $K$; we neglect this term from now on.

The kinetic term $\propto \pi_v^2$ comes with an overall
factor $\frac12 \ex^{2\alpha} = \frac12 (1+w)
(Q/\gamma)^{2s}$, which is exactly of the form obtained in
\eqref{T1}, showing that \eqref{Utransf} is indeed the
quantum analog of the classical canonical transformation
\eqref{ntoms} with $Z$ given by \eqref{Zs}. Indeed, the
wavelength term $\propto k^2$, upon factorizing $\frac12
(1+w) (Q/\gamma)^{2s}$ appearing in \eqref{T1} yields
exactly the required factor $(Q/\gamma)^{4(r_\F-s)} wk^2$ as
in \eqref{T2}.

The remaining terms read
$$ \left( \beta \ex^\alpha + \ex^\alpha \beta 
\right)^2 = \frac{s^2}{1+w} \left(\frac{Q}{\gamma}\right)^{-2s} 
\left[ -\frac14 (1+s)(5+3s) Q^{-4} + 2 i (1+s) Q^{-3} P + 
Q^{-2}P^2 \right],
$$
and
$$
2(\dot{\alpha}\beta+\beta\dot{\alpha}) =
\frac{s^2}{2(1+w)}\left(\frac{Q}{\gamma}\right)^{-2s}
\left[-(4s^2+8s+5)Q^{-4}+8i(1+s)Q^{-3}P+4Q^{-2}P^2\right],
$$
which can be directly evaluated using the above expressions,
as well as
$$
2 \dot\beta = \frac{s}{1+w}
\left(\frac{Q}{\gamma}\right)^{-2s} \left\{ \left[ 2 K +
\frac12 (1+s)(1+2s)(3+2s) \right] Q^{-4} - 2 i (1+s) (1+2s) 
Q^{-3} P - (1+2s) Q^{-2}P^2 \right\},
$$
obtained by calculating the commutator of $\beta$ with the
background Hamiltonian.

Finally, the total potential term can be written as
\begin{align}
\frac12(1+w) \left( \frac{Q}{\gamma} \right)^{2s}\left\{
\left( \frac{Q}{\gamma} \right)^{4r_\F-4s} w k^2+
\frac{s(s+1)}{(1+w)^2}
\left(\frac{Q}{\gamma}\right)^{-4s}Q^{-2} 
\left[P^2+2i(1+s)Q^{-1}P + g(s) Q^{-2}\right] \right\}v^2,
\label{AppFinal}
\end{align}
\end{widetext}
where
$$
g(s) = -\frac{8K+6+17s+16s^2+3s^3}{4(s+1)}.
$$
Comparison with the Hamiltonian \eqref{T2Q} reveals that the
canonical transformation provides an equivalent result
provided the choice $g_1(s) \to 1$ is made, and that $g_2(s)
\to g(s)$.

\bibliography{references}

\end{document}